\documentclass[english,10pt]{article}

\usepackage[T1]{fontenc}
\usepackage[latin1]{inputenc}
\usepackage{geometry}
\usepackage{float}
\usepackage{multirow}
\usepackage{mathrsfs}
\usepackage{amsmath}
\usepackage{amssymb}
\usepackage{graphicx}
\usepackage{hyperref}
\usepackage[color=yellow]{todonotes}
\hypersetup{
    colorlinks=true,
    linkcolor=blue,
    filecolor=magenta,      
    urlcolor=cyan,
    citecolor = blue,
}

\usepackage{yhmath}
\makeatletter

\usepackage{bm}
\numberwithin{equation}{section}

\usepackage{algorithm}
 \usepackage{algorithmicx}
\floatstyle{ruled}
\newfloat{algorithm}{tbp}{loa}
\providecommand{\algorithmname}{Algorithm}
\floatname{algorithm}{\protect\algorithmname}

\usepackage{algorithm}
\usepackage{algorithmicx}
\usepackage{amsthm}
\usepackage{mathrsfs}
\usepackage{amssymb}
\usepackage{amsfonts}
\usepackage{epsfig}
\usepackage{epstopdf}
\usepackage{bm}
\usepackage{mathrsfs}
\usepackage{enumerate}

\@ifundefined{definecolor}{\@ifundefined{definecolor}
 {\@ifundefined{definecolor}
 {\usepackage{color}}{}
}{}
}{}

\usepackage{subfig}\usepackage[all]{xy}

\newtheorem{rem}{Remark}[section]

\newcounter{hypA}

\newcounter{hypB}

\newcounter{hypD}

\newcommand{\dif}{\mathrm{d}}

\usepackage{babel}\date{}

\makeatother

\begin{document}

\begin{center}

{\Large \textbf{Multilevel Monte Carlo for a class of Partially Observed Processes in Neuroscience}}

\vspace{0.5cm}

BY  MOHAMED MAAMA$^{1}$, AJAY JASRA$^{2}$ \& KENGO KAMATANI$^{3}$   

{\footnotesize $^{1}$Applied Mathematics and Computational Science Program,  Computer, Electrical and Mathematical Sciences and Engineering Division, King Abdullah University of Science and Technology, Thuwal, 23955-6900, KSA.}\\
{\footnotesize $^{2}$School of Data Science, The Chinese University of Hong Kong, Shenzhen, CN.}\\
{\footnotesize $^{3}$Institute of Statistical Mathematics,  Tokyo 190-0014,  JP.}\\
{\footnotesize E-Mail:\,} \texttt{\emph{\footnotesize maama.mohamed@gmail.com, ajay.jasra@cuhk.edu.cn kamatani@ism.ac.jp}}

\end{center}

\begin{abstract}
In this paper we consider Bayesian parameter inference associated to a class of partially observed 
stochastic differential equations (SDE) driven by 
jump processes. Such type of models can be routinely found in applications, of which we focus upon the case of neuroscience. The data are assumed to be observed regularly in time and driven by the SDE model with unknown parameters. In practice the SDE
may not have an analytically tractable solution and this leads naturally to a time-discretization.
We adapt the multilevel Markov chain Monte Carlo method of \cite{jasra_bpe_sde}, which works with a hierarchy of time discretizations and show empirically and theoretically that this is preferable to using one single time discretization. The improvement is in terms of the computational cost needed to obtain a pre-specified numerical error. Our approach is illustrated on models applied to examples of Bayesian inference problems using both simulated and real observed data from neuroscience.
\\
\noindent \textbf{Keywords}: Multilevel Monte Carlo, Neuroscience Models, Stochastic Differential Equations, Action Potential Data, Bayesian Inference.
\end{abstract}

\section{Introduction}

The brain stands as one of the most complex systems known, containing billions of neurons. They are responsible for various functions, including memory, vision, motor skills, sensory perception, emotions, etc. These neurons facilitate the exchange of electrical signals through specialized junctions known as synapses. Broadly categorized into two types -electrical and chemical synapses- these fundamental connectors play important roles in the transmission of information within the brain. In electrical synapses, communication between neurons occurs directly. On the other hand, chemical synapses rely on neurotransmitters to convey messages. These neurotransmitters traverse the synaptic cleft, binding to receptors on the synaptic cell's membrane. Depending on their nature, neurotransmitters can either enhance or diminish the likelihood of generating an action potential in the postsynaptic neuron, giving rise to excitatory or inhibitory synapses, respectively.
One of the initial steps toward understanding certain areas of the brain and their functions involves grasping how simpler networks of neurons operate. This direction was first explored in the last century by Louis Lapicque in the 1900s, and Alan Lloyd Hodgkin and Andrew Fielding Huxley in the 1950s. 

Neuronal populations are invaluable as a fundamental model for understanding the intricate dynamics occurring in various regions of the brain, including the primary visual cortex in mammals (V1) \cite{Chariker2, Adi3}. The study of these large-scale networks is an expansive field of research, they have attracted a lot of attention from the applied science, computational mathematics, and statistical physics communities. The advantage of using the integrate-and-fire systems as a simplified model of neurons is that they are fast and efficient to simulate, see for example \cite{Adi1}. 
We are interested in learning several unknown parameters of leaky integrate-and-fire networks. Due to these numerical parameter regimes, many emergent behaviors in the brain are identified, such as synchronization, spike clustering or partial synchronization, background patterns, multiple firing events, gamma oscillations, and so on, see for example \cite{Chariker1,Chariker2,Maama,Adi3,Adi2} for more details.

\textcolor{black}{ Throughout the brain, neurons primarily communicate with each other through electrical impulses, commonly referred to as spikes.  Consequently, understanding the dynamics of membrane potential in an individual neuron is of utmost importance. Experimentally, intracellular recordings yield observations, typically measured at intervals of $0.1 \; ms$ (e.g. \cite{DataRat}). There is an increasing demand for robust methods to accurately estimate biophysically relevant parameters from such data (\cite{ DS, InferringS, NeuralEngineering, DataAssi}). Given that only the membrane potential can be measured, this complicates the statistical inference and parameter estimation process for these partially observed detailed models. Generally, parameters are estimated using Markov chain Monte Carlo algorithms. Based on the literature (e.g. \cite{DS}) and our expertise in computational statistics by developing algorithms for Bayesian parameter estimation, we are confident that our methodology is well-suited for application to experimental neuroscience data (e.g. \cite{DataRat}).}

This work deals with the inference, estimating unknown parameters, and numerical simulations of networks of leaky integrate-and-fire systems somewhere in the brain. The coupling mechanism depends on stochastic feedforward inputs, while the incorporation of recurrent inputs occurs through excitatory and inhibitory synaptic coupling terms. Furthermore, we applied our methodology to the stochastic Izhikevich model, using real action potential recordings data.

In this article we adopt the method of \cite{jasra_bpe_sde} to allow Bayesian parameter inference for a class of partially observed 
SDE models driven by jump processes. These type of models allow one to naturally fuse real data to well-known and relevant models in neuroscience. Simultaneously, such models can be rather complicated to estimate as, in the guise we consider them, they form a class of complex hidden Markov models, where the hidden process is an SDE. Naturally, such models are subject to several issues, including being able to estimate parameters and even being able to access the exact model. The approach that is adopted in this article is to first, as is typical in the literature, to time discretize the SDE and then to use efficient schemes on the approximate model. Our approach is to use Markov chain Monte Carlo (MCMC) combined with the popular multilevel Monte Carlo method of \cite{giles,hein}. This approach is able to reduce the computational cost to achieve a pre-specified mean square error. We provide details on the methodology combined with theoretically based guidelines on how to select simulation parameters to achieve the afore-mentioned cost reduction. This is illustrated on several neuroscience models.
We remark that we are not the first to consider parameter estimation in such contexts.
One example is the article of \cite{panin} whom consider related models but from a seemingly different perspective, in that latent process is observation driven and direct likelihood inference is the objective.  The model used for statistical inference is, ultimately quite different from that considered here and one would expect it to be used in different problems than our approach.

The structure of this paper is as follows. In Section \ref{sec:model} we give details on the model that is to be considered, whilst highlighting some applications in neuroscience. In Section \ref{sec:meth} we detail our computational methodology. In Section \ref{sec:numerics} we give our numerical results. In section \ref{sec:discussion} we provide a discussion part. Finally section \ref{sec:conclusion} is dedicated to presenting our conclusion.

\section{Model}\label{sec:model}

\subsection{SDE Model}\label{sec:sde_mod}

We consider a $d$-dimensional stochastic process $\{X_t\}_{t\in [0,T]}$ driven by a $m$-dimensional Poisson process 
$\{N_t\}_{t\in[0,T]}$ where each (one-dimensional) component $N_{j,t}$, $j\in\{1,\dots,m\}$ is
a one-dimensional Poisson process of rate $\lambda\in\mathbb{R}^+$
independent of all other Poisson processes.  A one-dimensional Poisson process is a non-negative,  integer valued collection of random variables, that starts at zero and increases by 1 at independent exponential times,  where the exponential random variable has a rate $\lambda$.  In other words,  one starts at zero and generates an exponential random variable of rate $\lambda$.  If that variable exceeds the time $T$ then the process remains at zero and one has simulated the process, otherwise at the exponential variable time the process jumps to 1, and one generates another exponential random variable and so on. For more precise details see \cite{daley}.
We will assume that $X_t$ solves the following stochastic differential equation: 
\begin{equation}\label{eq:rode}
\dif X_t = f_{\theta}(X_t,t)~\dif t + \Sigma_{\theta}~\dif N_t
\end{equation}
where $\theta\in\Theta$ is an unknown parameter,  $f:\Theta\times\mathbb{R}^d\times[0,T]\rightarrow\mathbb{R}^d$, $\Sigma:\Theta\rightarrow\mathbb{R}^{d\times m}$ (real $d\times m$ matrices) and $X_0=x_0^*\in\mathbb{R}^d$ is known. 
We note that the approach to be described can deal with a general point-process (see \cite{daley})
$\{\Xi_t\}_{t\in[0,T]}$ but we do not consider such processes in our simulations, hence this level of abstraction is omitted. 
$\theta$ is an unknown collection of static parameters for which we would like to infer on the basis of data. We assume that the \eqref{eq:rode} has a unique solution and we do not consider this aspect further as it is the case in all of our examples.

\begin{rem}\label{rem:path}
Let $C_{p}$ be the collection of real $d-$dimensional piecewise continuous functions 
and let $f:\Theta\times C_{p}\times[0,T]\rightarrow\mathbb{R}^d$ be a function.   
In many applications in neuroscience one often considers the path-dependent process
\begin{equation}
\dif X_t = f_{\theta}( \{X_s\}_{s\in [0,t]},t)~\dif t + \Sigma_{\theta}~\dif N_t.
\end{equation}
The methodology that we are to describe can easily be modified to this case with only notational changes and we will consider exactly these type of models in Section \ref{sec:coupled_neurons}; we do not however use this model in the forthcoming exposition for brevity. The methodology we describe does not change (up-to some details) but the theory of such processes is different to that of \eqref{eq:rode} and as such we do not claim that the forthcoming theoretical guidelines will work in such contexts, although empirical evidence in Section \ref{sec:coupled_neurons} suggest that they do.
\end{rem}

In many cases of practical interest, even though \eqref{eq:rode} has a unique solution, that may not be available in an analytically tractable manner. As a result, we assume that one has to time-discretize \eqref{eq:rode} appropriately with time step $\Delta_l=2^{-l}$. The discretization scheme is often very specific to the problem in hand and so we elaborate the discussion the case of a particular model, so as to make the discussion concrete.

\subsubsection{Example}

We consider a simple random dynamical system in neuroscience (see e.g.~\cite{neuron_model}), here $d=m$. We set
\begin{equation}
\dif X_t = f_{S^{\text{Q}}}(X_t,t)\dif t + S^{\text{dr}} \; \dif N_t, \quad \text{Q} \in \{\text{E,I}\}
\end{equation}
where $\theta=(S^{\text{Q}},S^{\text{dr}})^{\top}$ are scalars, $f_{S^{\text{Q}}}:\mathbb{R}^d\times[0,T]\rightarrow\mathbb{R}^d$ is left arbitrary and $\{N_t\}_{t\in[0,T]}$ is $d-$dimensional vector of independent homogeneous Poisson processes each of rate $\lambda\in\mathbb{R}^+$.


The discretization that we employ is the simple Euler method, as we now describe. We set for $k\in\{0,\dots,T\Delta_l^{-1}-1\}$:
\begin{equation}\label{eq:disc1}
X_{(k+1)\Delta_l}^l = X_{k\Delta_l}^l + \Delta_l \; f_{S^{\text{Q}}}(X_{k\Delta_l}^l,k\Delta_l) + S^{\text{dr}} \; \left(N_{(k+1)\Delta_l}-N_{k\Delta_l}\right).
\end{equation}
Note that this recursion can be simulated exactly,  so long as one has an initialization of the discretized process and one can evaluate the drift function $f$. The increments $N_{(k+1)\Delta_l}-N_{k\Delta_l}$ are simulated as described at the start of Section \ref{sec:sde_mod} independently of all other random variables.

In the context of the accuracy of the discretized recursion \eqref{eq:disc1} in terms of the true SDE \eqref{eq:rode}, one can use the strong error.  Informallly,  this is simply the expectation of the square of the Euclidean distance at a given time between the solution of \eqref{eq:rode} and the value at the same time produced by \eqref{eq:disc1}. Often these decay as $\mathcal{O}(\Delta_l^{\beta})$ and $\beta$ is called the strong error rate; naturally the larger the rate, the better the approximation.

As noted in \cite{bruti}, in general and under assumptions, the Euler method has a strong error rate of 1, however, in this case where the diffusion term is a constant $S^{dr}$ this should improve to 2 as it corresponds to the so-called `order 1.0 strong Taylor approximation' \cite[Section 3.5]{bruti}. These results are useful when we describe the approach we adopt for parameter estimation in Section \ref{sec:meth}.

\subsection{Data Model}

We consider observations, $Y_1,Y_2,\dots,Y_T$ that are seen at regular time points; for simplicity of exposition we suppose that these are unit times, but this need not be so.  We suppose that for each $k\in\{1,\dots,T\}$, $Y_k\in\mathsf{Y}$, where $\mathsf{Y}$ is some arbitrary space that we do not specify, but for instance one could have $\mathsf{Y}=\mathbb{R}$.
Denoting by $p$ as a generic probability density function we shall assume the following probabilistic structure:
\begin{equation}
p(y_{1:T}|x_{1:T}) = \prod_{k=1}^T g_{\phi}(y_k\mid x_k)
\end{equation}
where $y_{1:T}=(y_1,\dots,y_T)^{\top}$, $\phi\in\Phi$ are unknown parameters,  $g:\Phi\times\mathbb{R}^d\times\mathsf{Y}\rightarrow\mathbb{R}^+$ is a probability density in $y\in\mathsf{Y}$ for any $(\phi,x)\in\Phi\times\mathbb{R}^d$.   We use $\phi$ to denote the unknown parameters in the function $g$ and
recall that $\theta$ are the unknown parameters in the latent process defined in \eqref{eq:rode}.

\subsection{Posterior}
%
%

In Bayesian statistics (e.g.~\cite{bernardo}) statistical inference is made on the basis of a probability density on the unknown random variables.  Before any data are observed,  one needs to formulate a prior probability density on the unknown random variables and to specify a probabilty model on the data given these unknown random variables.  There is then a conditional probability mechanism called Bayes theorem that enables one to combine the data model with the prior to give a probability density on the unknown random variables; the posterior density.  Statistical inference,  such as parameter estimation can be based upon the posterior density,  for instance by computing posterior expectations.  In many cases of interest the latter cannot be computed but can be numerically approximated by sampling from the posterior; see \cite{monte} for an introduction to Bayesian computation.

In our case the unknown random variables are $x_{1:T}$ (recall the initial value is known i.e.~$x_0=x_0^*$) and the parameters $(\theta,\phi)$.  We have already formulated a model for 
$x_{1:T}$ via \eqref{eq:rode} and we let $\nu:=(\theta,\phi)^{\top}$ and $\pi_{\textrm{pr}}$ be a prior probability density on $\Theta\times\Phi$.
Let $p_{\theta}(x_t\mid x_{t-1})$ denote the Markov transition density of the process $\{X_t\}_{t\in[0,T]}$ (as the process is a L\'evy driven stochastic differential equation) which is assumed to exist.  If there is no transition density the posterior can be formulated using a measure theoretic approach which we do not, for simplicity.  The posterior probability density can be written
\begin{equation}\label{eq:post_cont}
\pi(\nu,x_{1:T}\mid y_{1:T}) \propto p(y_{1:T}\mid x_{1:T},\phi)\left\{\prod_{t=1}^T p_{\theta}(x_t\mid x_{t-1})
\right\}
\pi_{\textrm{pr}}(\nu).
\end{equation}
In practice, we simply work with the time discretized transition $p_{\theta}^l(x_t\mid x_{t-1})$ where $l$ is the index for time discretization that can be simulated (for instance) using 
\eqref{eq:disc1} and denote the resulting posterior $\pi^l$. The objective is now to sample from $\pi^l$ in order to approximate expectations w.r.t.~it.

\section{Computational Methodology}\label{sec:meth}

The computational approach that we use has been introduced in \cite{jasra_bpe_sde} (see also \cite{chada_ub,ml_rev,ml_fbm}) and we recapitulate for the context of the neuroscience models that we will consider.

\subsection{Markov Chain Monte Carlo}

In this section we introduce a method that can approximate expectations w.r.t.~$\pi^l$ for some fixed $l\in\mathbb{N}$.  \textcolor{black}{In essence,  this consists of running a discrete time ergodic Markov chain that will admit $\pi^l$ as its stationary distribution.  Then one can estimate expectations of functions w.r.t.~$\pi^l$ by simply computing the average of the function values at each point of the simulated chain.  A comprehensive introduction can be found in \cite{monte}.}

\textcolor{black}{To define the algorithm (which is Algorithm \ref{alg:mcmc}),  we first need a preliminary method that is described in Algorithm \ref{alg:pf}, which is run with $l$ and $\nu$ fixed and known.
In the context of sampling from $\pi^l$ the objective is to provide an unbiased estimator of
the marginal likelihood
\begin{equation}
p_{\nu}^l(y_{1:T}) = \int_{\mathbb{R}^{dT}}p(y_{1:T}\mid x_{1:T},\phi)\left\{\prod_{t=1}^T p_{\theta}^l(x_t\mid x_{t-1})\right\}
\dif x_1\dots \dif x_T.
\end{equation}}
The estimator has been extensively studied in the literature,  such as a proof of unbiasedness; see \cite{delm_book} for example.  \textcolor{black}{The reason why such an estimator is needed,  is simply that even though
$p_{\nu}^l(y_{1:T})$ is not available analytically,  it is enough that one can obtain an unbiased estimate of it, so
as to sample from $\pi^L$; see \cite{andrieu}.
Algorithm \ref{alg:pf} has been originally derived for performing sequential filtering and marginal likelihood estimation; see e.g.~\cite{delm_book} and the references therein.  It is called particle simulation and consists of simulating $S$ samples,  sequentially and in parallel.  It begins by sampling $S\in\mathbb{N}$ samples from the dynamics
$p_{\theta}^l(~\cdot~\mid x_{0}^*)$ which can be done by Euler discretization as is employed in \eqref{eq:disc1}.
The $S$ samples then undergo a resampling mechanism,  to correctly incorporate the data into the approximation of 
distribution proportional to $g_{\theta}(y_1|x_1)p_{\theta}^l(x_1|x_0^*)$. This occurs by sampling with replacement using a weight function associated to $g_{\theta}$ (see Algorithm \ref{alg:pf}). This produces $S$ samples and then one moves forward in time as described in Algorithm \ref{alg:pf},  which along the way has a mechanism to approximate the marginal likelihood. The choice of $S$ is described below.
}
In Algorithm \ref{alg:pf}, one returns a trajectory $X_{1:T}\in\mathbb{R}^{dT}$ and an estimator of 
$p_{\nu}^l(y_{1:T})$.  

The main MCMC method is given in Algorithm \ref{alg:mcmc}.  \textcolor{black}{
The method requires the ability to sample from the prior $\pi_{\textrm{pr}}$ which is often simple,
as the associated distribution is often selected as a something well-known (e.g.~a Gamma distribution)
and a proposal distribution $q$ on the parameter space.  The choice of the latter has been extensively investigated in the literature and we refer the reader to \cite{monte} and the references therein.}
In Algorithm \ref{alg:mcmc} for any functional $\varphi:\Theta\times\Phi\times\mathbb{R}^{dT}\rightarrow\mathbb{R}$
that is integrable w.r.t.~the posterior $\pi^l$ the expectation w.r.t.~$\pi^l$ can
be estimated using:
\begin{equation}
\widehat{\pi}^l(\varphi)^M := \frac{1}{M+1}\sum_{k=0}^M\varphi(\nu(k),X_{1:T}^l(k)).
\end{equation}
It is known that the estimator converges and finite sample convergence rates are also known; see \cite{chada_ub,jasra_bpe_sde,jasra_cont,ml_rev} and the references therein. The choice of $S$ is often $S=\mathcal{O}(T)$
and we do this in our numerical simulations.

\begin{algorithm}
\begin{enumerate}
\item{Input: $(\nu,S,T,l)\in(\Theta\times\Phi)\times\mathbb{N}^3$. Set $k=1$ and $x_0^i=x_0^*$, $i\in\{1,\dots,S\}$. Set $\hat{p}^l(y_{1:0})=1$}
\item{Sampling: For $i\in\{1,\dots,S\}$ sample $X_k^i\mid X_{k-1}^i$ from the discretized dynamics $p_{\theta}^l(~\cdot~\mid x_{k-1}^i)$. If $k=T$ go to step 4.~otherwise go to step 3..}
\item{Resampling: For $i\in\{1,\dots,S\}$ compute $W_k^i=g_{\phi}(y_k\mid x_k^i)/\{\sum_{j=1}^Sg_{\phi}(y_k\mid x_k^j)\}$. 
Set $\hat{p}^l(y_{1:k}) = \hat{p}^l(y_{1:k-1})\tfrac{1}{S}\sum_{j=1}^Sg_{\phi}(y_k\mid x_k^i)$.
Sample $S$ times with replacement from $(X_{1:k}^1,\dots,X_{1:k}^S)$ using the probability mass function $W_k^{1:S}$ and denote the resulting samples $(X_{1:k}^1,\dots,X_{1:k}^S)$.  Set $k=k+1$ and go to the start of step 2..}
\item{Select Trajectory:  For $i\in\{1,\dots,S\}$ compute $W_T^i=g_{\phi}(y_T\mid x_T^i)/\{\sum_{j=1}^Sg_{\phi}(y_T\mid x_T^j)\}$ and
select one trajectory $X_{1:T}^i$ from $(X_{1:T}^1,\dots,X_{1:T}^S)$ by sampling once from the probability mass function $W_T^{1:S}$. 
Set $\hat{p}^l(y_{1:T}) = \hat{p}^l(y_{1:T-1})\tfrac{1}{S}\sum_{j=1}^Sg_{\phi}(y_T\mid x_T^j)$.
Go to step 5..}
\item{Output: Selected trajectory and $\hat{p}^l(y_{1:T})$.}
\end{enumerate}
\caption{Particle Filter}
\label{alg:pf}
\end{algorithm}

\begin{algorithm}
\begin{enumerate}
\item{Input: $(M,S,T,l)\in\mathbb{N}^4$, the number of MCMC iterations, samples in Algorithm \ref{alg:pf}, time steps in the model and level $l$. Also specify a positive Markov transition density $q$ on $\Theta\times\Phi$.}
\item{Initialize: Sample $\nu(0)$ from the prior $\pi_{\textrm{pr}}$ and then run Algorithm \ref{alg:pf} with parameter $\nu(0)$ denoting the resulting trajectory as $X_{1:T}^l(0)$ and $\hat{p}^{l,0}(y_{1:T})$ the marginal likelihood estimator. Set $k=1$ and go to step 3..}
\item{Iterate: Generate $\nu'|\nu(k-1)$ using the Markov transition $q$. Run Algorithm \ref{alg:pf} with parameter $\nu'$ and denote the output $(X_{1:T}',\hat{p}^{l,'}(y_{1:T}))$. Sample $U\sim\mathcal{U}_{[0,1]}$ (the uniform distribution on $[0,1]$) if
$$
U < \min\left\{1,\frac{\hat{p}^{l,'}(y_{1:T})~~~~\pi_{\textrm{pr}}(\nu')~~~~~q(\nu(k-1)\mid \nu')}{\hat{p}^{l,k-1}(y_{1:T})\pi_{\textrm{pr}}(\nu(k-1))q(\nu'\mid \nu(k-1))}\right\}
$$
then set $(\nu(k),X_{1:T}^{l}(k),\hat{p}^{l,k}(y_{1:T}))=(\nu',X_{1:T}',\hat{p}^{l,'}(y_{1:T}))$. Otherwise set
$(\nu(k),X_{1:T}^{l}(k),\hat{p}^{l,k}(y_{1:T}))=(\nu(k-1),X_{1:T}^{l}(k-1),\hat{p}^{l,k-1}(y_{1:T}))$.
If $k=M$ go to step 4., otherwise set $k=k+1$ and return to the start of step 3..}
\item{Output: $(\nu(0:M),X_{1:T}^{l}(0:M))$.}
\end{enumerate}
\caption{Markov chain Monte Carlo}
\label{alg:mcmc}
\end{algorithm}

\subsection{Multilevel Markov Chain Monte Carlo}

\subsubsection{Multilevel Identity}

We now introduce the MLMCMC method that was used in \cite{jasra_bpe_sde}, which is based upon the multilevel Monte Carlo method \cite{giles,hein}. The basic notion is as follows, first define for $\varphi:\Theta\times\Phi\times\mathbb{R}^{dT}\rightarrow\mathbb{R}$:\textcolor{black}{
\begin{equation}
\pi^l(\varphi) := \int_{\Theta\times\Phi\times\mathbb{R}^{dT}}\varphi(\nu,x_{1:T})\pi^l(\nu,x_{1:T}\mid y_{1:T})
\dif\nu\dif x_{1}\dots\dif x_T
\end{equation}}
where we assume that the R.H.S.~is finite. Then we have the following collapsing sum representation for any $L\geq 2$
\begin{equation}
\pi^L(\varphi) = \pi^1(\varphi) + \sum_{l=2}^L \left\{\pi^l(\varphi)-\pi^{l-1}(\varphi)\right\}.
\end{equation}
To approximate the R.H.S.~of this identity, we already have a method to deal with $\pi^1(\varphi)$ in Algorithm \ref{alg:mcmc}.
To deal with $\pi^l(\varphi)-\pi^{l-1}(\varphi)$ as has been noted in many articles e.g.~\cite{giles,ml_rev} it is not sensible to approximate $\pi^l(\varphi)$ and $\pi^{l-1}(\varphi)$ using statistically independent methods. 
\textcolor{black}{To understand this point,  essentially employing Algorithm \ref{alg:mcmc} to approximate
$\pi^l(\varphi)$ and then independently $\pi^{l-1}(\varphi)$ does not take advantage of the point that
as $l$ grows,  these expectations become increasingly similar and, indeed, the difference will converge to zero
as $l$ increases.  The main crux of MLMCMC methods is to exactly leverage this similarity by using a mechanism called coupling (described below).  This latter method, if sufficiently well employed can produce a variance reduction versus simply approximating $\pi^L(\varphi)$ on its own and ultimately lead to the cost of computation being reduced. This observation in the context of MCMC is hardly trivial and significant efforts have been made to realize this approach; see e.g.~\cite{ml_rev} and the references therein.}
We give an identity which has been used in \cite{jasra_bpe_sde} to provide a useful (in a sense to be made precise below) approximation of $\pi^l(\varphi)-\pi^{l-1}(\varphi)$.

\subsubsection{Multilevel Identity of \cite{jasra_bpe_sde}}
\textcolor{black}{
We begin by introducing the notion of a coupling of $p_{\theta}^l(x_t\mid x_{t-1})$ and $p_{\theta}^{l-1}(x_t'\mid x_{t-1}')$, $l\geq 2$, $T\geq t\geq 1$.
Set $z_t=(x_t,x_t')$, by a coupling we mean any probability kernel of the type $\check{p}^l(z_t\mid z_{t-1})$ such that for
any set $A\subseteq\mathbb{R}^d$, $z_{t-1}\in\mathbb{R}^{2d}$
\begin{equation}\label{eq:coup}
\int_{A\times\mathbb{R}^d} \check{p}^l(z_t\mid z_{t-1})\dif z_t = \int_Ap_{\theta}^l(x_t\mid x_{t-1})\dif x_t \quad \textrm{and} \quad
\int_{\mathbb{R}^d\times A} \check{p}^l(z_t\mid z_{t-1})\dif z_t = \int_Ap_{\theta}^{l-1}(x_t'\mid x_{t-1}')\dif x_t.
\end{equation}
Such a coupling always exists in our case for instance:
\begin{equation}
\check{p}^l(z_t\mid z_{t-1}) = p_{\theta}^l(x_t\mid x_{t-1})p_{\theta}^{l-1}(x_t'\mid x_{t-1}')
\end{equation}
and we detail a more useful one below.}

In the context of the Euler discretization \eqref{eq:disc1} such a coupling is easily constructed via the synchronous coupling as we now describe.
\begin{enumerate}
\item{Input; starting point $z_{t-1}$ and level $l\geq 2$.}
\item{Generate $N_{t-1+\Delta_l}-N_{t-1},\dots,N_{t}-N_{t-\Delta_l}$ as i.i.d.~$\mathcal{P}_d(\lambda\Delta_l)$ ($d-$independent Poisson random variables each of rate $\lambda\Delta_l$) random variables.}
\item{Compute $N_{t-1+\Delta_{l-1}}-N_{t-1},\dots,N_{t}-N_{t-\Delta_{l-1}}$ by summing the relevant consecutive increments from step 2..}
\item{Run the recursion \eqref{eq:disc1} at levels $l$ and $l-1$ using the starting points $x_{t-1}$ and $x_{t-1}'$ respectively, with the increments from step 2.~and step 3.~respectively, up-to time $t$.}
\item{Output: $z_t$.}
\end{enumerate}

The approach of \cite{jasra_bpe_sde} is now to introduce a probability density of the type:
\begin{equation}\label{eq:coup_post}
\check{\pi}^l(\nu,x_{1:T}^l,x_{1:T}^{l-1}) \propto \left\{\prod_{k=1}^T \check{g}_{\phi,k}(z_k^l)~\check{p}^l( z_k^l\mid z_{k-1}^{l})\right\}\pi_{\textrm{pr}}(\nu)
\end{equation}
where we use the notation $z_k^l=(x_k^l,x_{k}^{l-1})$ and $\check{g}_{\phi,k}(z_k^l)$ is an arbitrary positive function; for instance $\check{g}_{\phi,k}(x_k^l,x_{k}^{l-1}) = \max\{g_{\phi}(y_k\mid x_k^l),g_{\phi}(y_k\mid x_k^{l-1})\}$ or 
$\check{g}_{\phi,k}(x_k^l,x_{k}^{l-1}) = \tfrac{1}{2}\{g_{\phi}(y_k\mid x_k^l)+g_{\phi}(y_k\mid x_k^{l-1})\}$.
\textcolor{black}{The motivation of considering such a probability density is perhaps far from obvious and
we try to explain why this is used in words.  The problem of simulating from a coupling of (in the sense of \eqref{eq:coup}) $\pi^l$ and $\pi^{l-1}$ is very challenging, however,  as described above simulating the coupling $\check{p}^l$ is relatively easy.  Thus the idea is then to use $\check{p}^l$ and then create an \emph{artificial} probability density to simulate from which is \emph{not} a coupling of $\pi^l$ and $\pi^{l-1}$.  Due to the latter fact,  we then have to use an appropriate correction in order to approximate the difference $\pi^l(\varphi)-\pi^{l-1}(\varphi)$ and this is done below.  The expectation is that the coupling  $\check{p}^l$ is strong enough that even though one needs a correction, the variance reduction that was discussed above can be achieved and that there is a genuine benefit of using the multilevel method through a cost reduction.  These features are now resolved in the sequel.} 

Set $z_{1:T}^l=(x_{1:T}^l,x_{1:T}^{l-1})$ and 
\begin{eqnarray}
R_{T,1}(z_{1:T}^l,\phi) & = & \prod_{k=1}^T\frac{g_{\phi}(y_k\mid x_k^l)}{\check{g}_{\phi,k}(x_k^l,x_{k}^{l-1})} \\
R_{T,2}(z_{1:T}^l,\phi) & = & \prod_{k=1}^T\frac{g_{\phi}(y_k\mid x_k^{l-1})}{\check{g}_{\phi,k}(x_k^l,x_{k}^{l-1})}.
\end{eqnarray}
\cite{jasra_bpe_sde} show that
\begin{eqnarray}
\pi^l(\varphi)-\pi^{l-1}(\varphi) & =
 \frac{\int_{\Theta\times\Phi\times\mathbb{R}^{2dT}}
\varphi(\nu,x_{1:T}^l)R_{T,1}(z_{1:T}^l,\phi)\check{\pi}^l(\nu,z_{1:T}^l)\dif\nu\dif z_1^l\dots\dif z_T^l 
}{\int_{\Theta\times\Phi\times\mathbb{R}^{2dT}}
R_{T,1}(z_{1:T}^l,\phi)\check{\pi}^l(\nu,z_{1:T}^l)\dif\nu\dif z_1^l\dots\dif z_T^l } -\nonumber \\
&
\frac{\int_{\Theta\times\Phi\times\mathbb{R}^{2dT}}
\varphi(\nu,x_{1:T}^{l-1})R_{T,2}(z_{1:T}^l,\phi)\check{\pi}^l(\nu,z_{1:T}^l)\dif\nu\dif z_1^l\dots\dif z_T^l 
}{\int_{\Theta\times\Phi\times\mathbb{R}^{2dT}}
R_{T,2}(z_{1:T}^l,\phi)\check{\pi}^l(\nu,z_{1:T}^l)\dif\nu\dif z_1^l\dots\dif z_T^l }.
\end{eqnarray}
The strategy that is employed then, is simply to produce an algorithm to approximate expectations w.r.t.~$\check{\pi}^l$, $l\in\{2,\dots,L\}$ and to run these algorithms independently. In the next section we give an MCMC method to approximate expectations w.r.t.~$\check{\pi}^l$.

\subsubsection{Multilevel MCMC Algorithm}

\textcolor{black}{To define an algorithm to sample from $\check{\pi}^l$ one should note that first
the probabilistic structure of \eqref{eq:coup_post} and \eqref{eq:post_cont} are very similar
(and note that $\pi^l$ also has this structure) except that the former has a larger state-space than the latter.
Hence,  to sample from $\check{\pi}^l$ one can use a very similar strategy that was used to sample from 
$\pi^l$, except that it is appropriately modified to the correct context.  This is what we do now.
}
To define the algorithm, we first need a method that is described in Algorithm \ref{alg:delta_pf} and is simply an analogue of 
Algorithm \ref{alg:pf}. The MCMC method to sample $\check{\pi}^l$ is then given in Algorithm \ref{alg:mlmcmc}. The MCMC method is simply a mild adaptation of Algorithm \ref{alg:mcmc} in the context of a different target distribution. The differences in the main are essentially notational.

In order to approximate $\pi^l(\varphi)-\pi^{l-1}(\varphi)$ one has the following estimator:
$$
\widehat{\pi^l(\varphi)-\pi^{l-1}(\varphi)}^M :=
$$
\begin{equation}
\frac{\tfrac{1}{M+1}
\sum_{k=0}^M
\varphi(\nu(k),x_{1:T}^l(k))R_{T,1}(z_{1:T}^l(k),\phi(k))
}{\tfrac{1}{M+1}
\sum_{k=0}^MR_{T,1}(z_{1:T}^l(k),\phi(k))} - 
\frac{\tfrac{1}{M+1}
\sum_{k=0}^M
\varphi(\nu(k),x_{1:T}^{l-1}(k))R_{T,2}(z_{1:T}^l(k),\phi(k))
}{\tfrac{1}{M+1}
\sum_{k=0}^MR_{T,2}(z_{1:T}^l(k),\phi(k))}.
\end{equation}
There have been many results proved about such estimators; see \cite{chada_ub,jasra_bpe_sde,jasra_cont,ml_rev}.

\begin{algorithm}
\begin{enumerate}
\item{Input: $(\nu,S,T,l)\in(\Theta\times\Phi)\times\mathbb{N}^3$. Set $k=1$ and $z_0^i=(x_0^*,x_0^*)$, $i\in\{1,\dots,S\}$. Set $\tilde{p}^l(y_{1:0})=1$}
\item{Sampling: For $i\in\{1,\dots,S\}$ sample $Z_k^i\mid Z_{k-1}^i$ from the coupled discretized dynamics $\check{p}_{\theta}^l(~\cdot~\mid z_{k-1}^i)$. If $k=T$ go to step 4.~otherwise goto step 3..}
\item{Resampling: For $i\in\{1,\dots,S\}$ compute $\check{W}_k^i=\check{g}_{\phi,k}(z_k^i)/\{\sum_{j=1}^S\check{g}_{\phi,k}(z_k^i)\}$. 
Set $\tilde{p}^l(y_{1:k}) = \tilde{p}^l(y_{1:k-1})\tfrac{1}{S}\sum_{j=1}^S\check{g}_{\phi,k}(z_k^i)$.
Sample $S$ times with replacement from $(Z_{1:k}^1,\dots,Z_{1:k}^S)$ using the probability mass function $\check{W}_k^{1:S}$ and denote the resulting samples $(Z_{1:k}^1,\dots,Z_{1:k}^S)$.  Set $k=k+1$ and go to the start of step 2..}
\item{Select Trajectory:  For $i\in\{1,\dots,S\}$ compute $\check{W}_T^i=\check{g}_{\phi,k}(z_T^i)/\{\sum_{j=1}^S\check{g}_{\phi,k}(z_T^i)\}$ and
select one trajectory $Z_{1:T}^i$ from $(Z_{1:T}^1,\dots,Z_{1:T}^S)$ by sampling once from the probability mass function $\check{W}_T^{1:S}$. 
Set $\tilde{p}^l(y_{1:T}) = \tilde{p}^l(y_{1:T-1})\tfrac{1}{S}\sum_{j=1}^S\check{g}_{\phi,k}(z_T^i)$.
Go to step 5..}
\item{Output: Selected trajectory and $\tilde{p}^l(y_{1:T})$.}
\end{enumerate}
\caption{Delta Particle Filter}
\label{alg:delta_pf}
\end{algorithm}

\begin{algorithm}
\begin{enumerate}
\item{Input: $(M,S,T,l)\in\mathbb{N}^3\times\{2,3,\dots\}$, the number of MCMC iterations, samples in Algorithm \ref{alg:pf}, time steps in the model and level $l$. Also specify a positive Markov transition density $q$ on $\Theta\times\Phi$.}
\item{Initialize: Sample $\nu(0)$ from the prior $\pi_{\textrm{pr}}$ and then run Algorithm \ref{alg:delta_pf} with parameter $\nu(0)$ denoting the resulting trajectory as $Z_{1:T}^l(0)$ and $\tilde{p}^{l,0}(y_{1:T})$ the normalizing estimator. Set $k=1$ and go to step 3..}
\item{Iterate: Generate $\nu'\mid \nu(k-1)$ using the Markov transition $q$. Run Algorithm \ref{alg:delta_pf} with parameter $\nu'$ and denote the output $(Z_{1:T}',\tilde{p}^{l,'}(y_{1:T}))$. Sample $U\sim\mathcal{U}_{[0,1]}$ if
$$
U < \min\left\{1,\frac{\tilde{p}^{l,'}(y_{1:T})~~~~~~\pi_{\textrm{pr}}(\nu')~~~~~~~q(\nu(k-1)\mid \nu')}{\tilde{p}^{l,k-1}(y_{1:T})~\pi_{\textrm{pr}}(\nu(k-1))~q(\nu'\mid \nu(k-1))}\right\}
$$
then set $(\nu(k),Z_{1:T}^{l}(k),\tilde{p}^{l,k}(y_{1:T}))=(\nu',Z_{1:T}',\tilde{p}^{l,'}(y_{1:T}))$. Otherwise set
$(\nu(k),Z_{1:T}^{l}(k),\tilde{p}^{l,k}(y_{1:T}))=(\nu(k-1),Z_{1:T}^{l}(k-1),\tilde{p}^{l,k-1}(y_{1:T}))$.
If $k=M$ go to step 4., otherwise set $k=k+1$ and return to the start of step 3..}
\item{Output: $(\nu(0:M),Z_{1:T}^{l}(0:M))$.}
\end{enumerate}
\caption{Bilevel Markov chain Monte Carlo}
\label{alg:mlmcmc}
\end{algorithm}

\subsubsection{Final Method and Estimator}

The approach that we use is then as follows:
\begin{enumerate}
\item{Choose $L$ the final level and $M_1,\dots,M_L$ the number of samples to be used at each level.}
\item{Run Algorithm \ref{alg:mcmc} at level 1 for $M_1$ iterations.}
\item{For $l\in\{2,\dots,L\}$, independently of step 2.~and all other steps run Algorithm \ref{alg:mlmcmc} for $M_l$ steps.}
\item{Return the estimator for any $\varphi:\Theta\times\Phi\times\mathbb{R}^{dT}\rightarrow\mathbb{R}$ that is (appropriately) integrable:
\begin{equation}
\widehat{\pi_L(\varphi)} = \widehat{\pi}^l(\varphi)^{M_1} + \sum_{l=2}^L \widehat{\pi^l(\varphi)-\pi^{l-1}(\varphi)}^{M_l}
\end{equation}
}
\end{enumerate}
The main issue is then how one can choose $L$ and $M_1,\dots,M_L$.

\subsubsection{Theoretical Guidelines}

To choose $L$ and $M_1,\dots,M_L$, one can appeal to the extensive theory in \cite{chada_ub,jasra_bpe_sde,jasra_cont,ml_rev} in the context of Euler discretization that was mentioned - we refer the reader to those papers for a more precise description of the mathematical results we will allude to. We do not give explicit proofs, but essentially the results that we suggest here can be obtained by a combination of the afore-mentioned works
and the results that are detailed in \cite{bruti} which suggests that the strong error rate is 2. Let $\epsilon>0$ be given,
one can choose $L$ so that $\Delta_L^2=\mathcal{O}(\epsilon^2)$; this comes from bias results on Euler discretizations and the work in \cite{jasra_bpe_sde} for diffusion processes. As the strong error rate is 2, one can choose $M_l=\mathcal{O}(\epsilon^{-2}\Delta_l^{3/2})$. These choices would give a mean square error of $\mathcal{O}(\epsilon^2)$ and the cost to achieve this is the optimal $\mathcal{O}(\epsilon^{-2})$. Using a single level, $L$ (i.e.~in the case of just using Algorithm \ref{alg:mcmc}) one would obtain the same mean square error at a cost of $\mathcal{O}(\epsilon^{-3})$.

\section{Numerical Results}\label{sec:numerics}

\subsection{\textcolor{black}{Integrate-and-Fire type models}}

We consider a network of excitatory and inhibitory neurons governed by the integrate-and-fire model.
\begin{figure}[H]
\centering
\includegraphics[width=9cm,height=3cm]{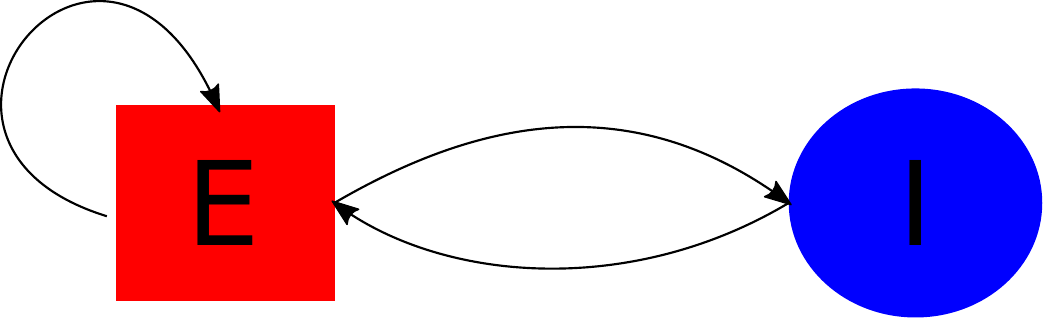}
 \caption{Excitatory-Inhibitory Neuronal Networks Scheme.}
    \label{fig:EI}
\end{figure}

The system of equations associated with our network is by adding a term of the excitatory and the inhibitory postsynaptic terms to the right-hand side of a single neuron $j$, which is given by the model:
\begin{equation}
 \dif V_{j,t} =  \left( \frac{1}{\tau_{V}} (V_{\text{reset}} - V_{j,t})  +  g^{\textrm{E}}(t)  -  g^{\textrm{I}}(t)  \right) \dif t + S^{\textrm{dr}} \dif N_{j,t} ,
\label{moneq2}
\end{equation}

\textcolor{black}{
Let $\mathbb{I}_{[0,\infty)}(x)$ be the indicator that $x$ is non-negative,  that is, 1 if $x\geq 0$ and $0$ otherwise.
 In this article, the excitatory synapse \ $g^{\textrm{E}}(t)$\ for a postsynaptic neuron\ $j$\ is governed by the equation:
\begin{equation}
  g^{\textrm{E}}(t) = \sum_{i \in \Gamma^{\textrm{E}} (j)}  S_{ij}^{\textrm{E}} \, \sum_{k^{'}=1}^{n_{i,t}^\textrm{E}} \mathbb{I}_{[0,\infty)}(t - t_{i,k^{'}}^{\textrm{syn,E}}). 
	\label{moneq3}
\end{equation}
and, the inhibitory synapse\ $g^{\textrm{I}}(t)$\ for a neuron\ $j$\ obeys 
\begin{equation}
  g^{\textrm{I}}(t) = \sum_{i \in \Gamma^{\textrm{I}} (j)}  S_{ij}^{\textrm{I}}\, \sum_{k^{'}=1}^{n_{i,t}^\textrm{I}} 
\mathbb{I}_{[0,\infty)}(t - t_{i,k^{'}}^{\textrm{syn,I}}). 
	\label{moneq4}
\end{equation} }
The notations will be further explained in the following paragraph.
These two formulations model the cumulative excitation and inhibition experienced by neuron $j$ within the network due to all action potential events that have occurred up to time $t$. We say a spike or an action potential at time\ $t$\ occurs if\ $V_{j,t}$\ of the presynaptic neuron crossed a threshold\ $V_{\text{thr}}$\ at that time and resets to $V_\text{{reset}}$, where $V_{j,t} \in [V_\text{{reset}} , V_{\text{thr}}] = [0,1]$. The notation $\Gamma^{\textrm{Q}} (j)$ denotes the set of the presynaptic $\text{Q}-$neurons of the neuron $j$, where $\text{Q} \in \left\{\text{E,I}\right\}$. A fundamental aspect of our networks is that our synapses are instantaneous. In formulation~\eqref{moneq3} above, the sequence time\ $\left\{t_{i,k^{'}}^{\textrm{syn,E}}\right\}_{k^{'}}^{}$\, where $n_{i,t}^\textrm{E}$ is the number up-to time $t$, are the times at which a kick from one of the excitatory neurons in the network is received by neuron\ $j$. Similarly, in ~\eqref{moneq4} the instants\ $\left\{t_{i,k^{'}}^{\textrm{syn,I}}\right\}_{k^{'}}^{}$\ (again where $n_{i,t}^\textrm{I}$ is the number up-to time $t$) are the times at which inhibitory kicks are received. The adjacency matrices $(S_{ij}^{E,I})$ represent the strengths of $E$ or $I$ synapses from neuron $i$ to neuron $j$ and map the network topology. Our network is driven by independent stochastic inputs at the right-hand side of equation~\eqref{moneq2}. By definition, we set $ S^{\text{dr}} \dif N_{j,t} = S^{\text{dr}} \sum_{k}^{} \mathbb{I}_{[0,\infty)} (\dif t - T_{j,k}^{\text{dr}}) $. Indeed, we use spike kicks from a Poisson process with rate\ $\lambda$ ($T_{j,k}^{\text{dr}}$ are the event times). The membrane leakage timescale $\tau_V  = 20 \; ms$, the feed-forward arrival times input to the neuron\ $j$\ are denoted by\ $\left\{T_{j,k}^{\text{dr}}\right\}_{k}^{}$, and\ $S^{\text{dr}}$\ is an external strength constant.

We are initiating our numerical simulations with a randomly generated network topology comprising $N_E$ excitatory neurons and $N_I$ inhibitory neurons, in which every neuron within the graph receives an independent external Poisson process. 
The Raster plots figure below illustrates how variations in parameters give rise to some emergent properties within the network.
For simplicity, we employ two straightforward special cases for the numerical simulations.

\begin{figure}[H]
\centering
\subfloat{\includegraphics[width=12cm,height=4.5cm]{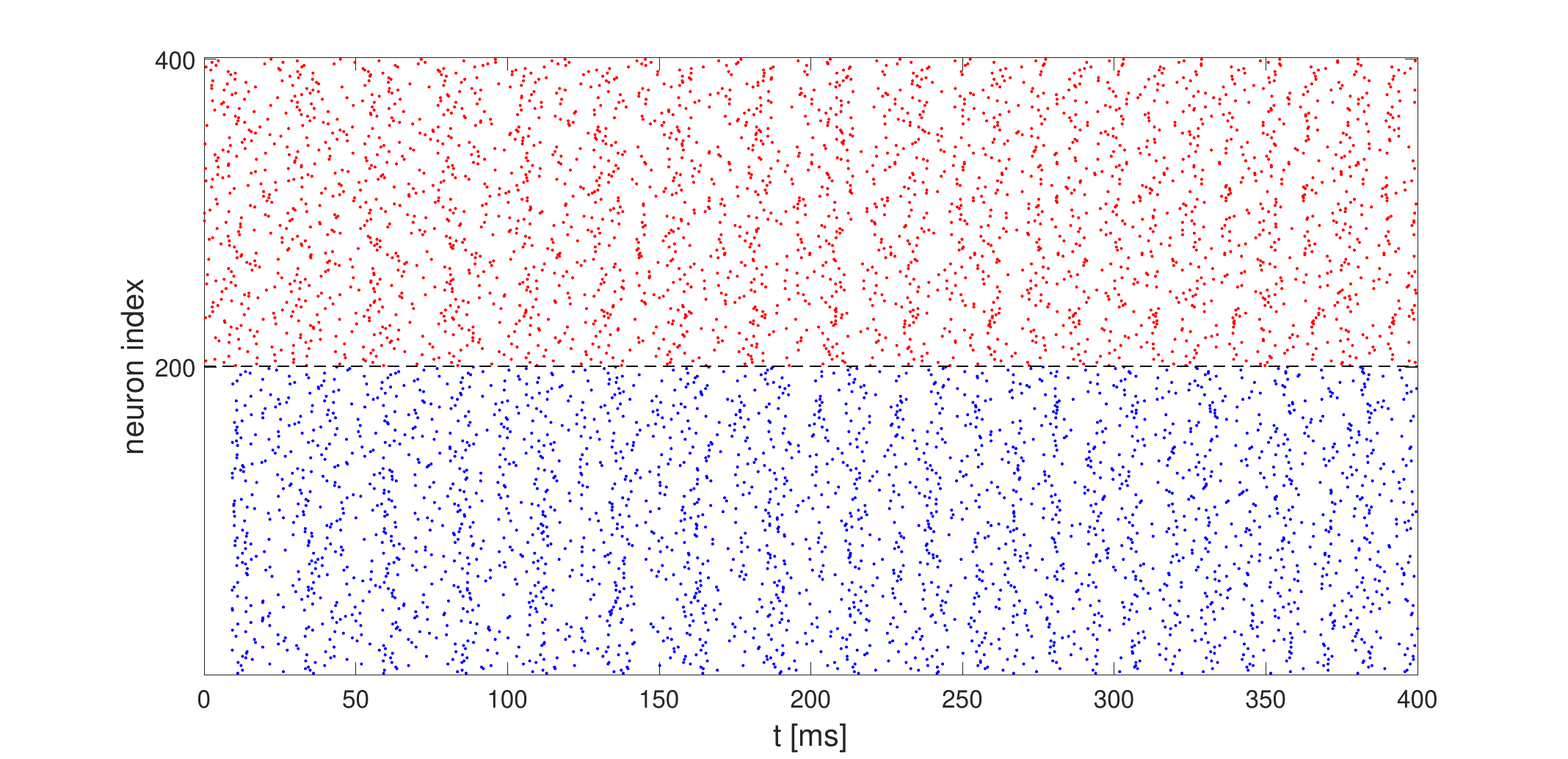}} \\
\subfloat{\includegraphics[width=12cm,height=4.5cm]{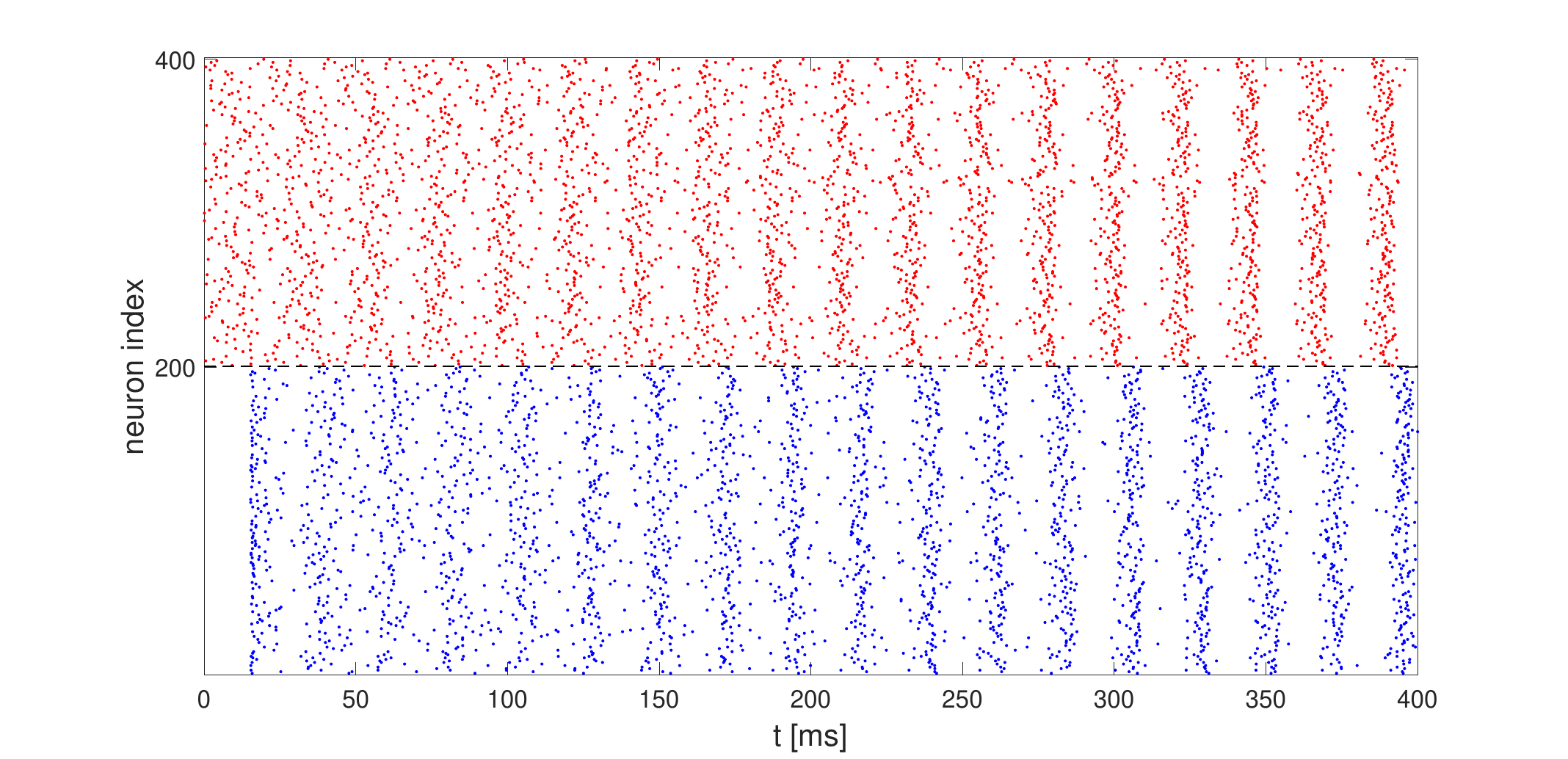}}
\caption{Raster plots of firing-activity for two different systems during $400 \; ms$, inhibitory (blue - bottom half) and excitatory (red - top half) integrate-and-fire neurons showing two different regimes. Top: A homogeneous property with minimal to no correlations among spike times., $ S^{\text{EE}} = S^{\text{II}} = S^{\text{EI}} = S^{\text{IE}} = 0.0028$. Bottom: A more synchronous regime with the majority of the network firing in perfect synchronization, $ S^{\text{EE}} = S^{\text{II}} = 0.009$, $ S^{\text{EI}} = S^{\text{IE}} = 0.007$. Each regime has network parameters, $N_{\text{E}} = N_{\text{I}} = 200$, and the Poisson random inputs are the amplitude $S^{\text{dr}} = 0.065$, the frequency per ms $\lambda = 0.55$. }
\label{fig:R}
\end{figure}

\subsubsection{Case 1: Integrate-and-Fire Model Driven by Poisson Spike Trains}

This first simple model is intentionally straightforward, featuring an integrate-and-fire mechanism of spiking neurons with independent random input injections, devoid of any coupling from other neurons in the network. It consists only of a simple SDE shot noise, described as follows:

\begin{equation}
 \dif V_t=  \Big(  \frac{1}{\tau_{V}} (V_{\text{reset}} - V_t)  \Big) \dif t + S^{\text{dr}}\dif N_t ,
\label{moneqIF}
\end{equation}

For our numerical experiments, with model \eqref{moneqIF}, we consider the prior for our parameter of interest as $ \theta=S^{\mathrm{dr}} \sim \mathcal{G}(0.01,0.005)$ where $\mathcal{G}(a,b)$ is the Gamma distribution with shape $a$ and scale $b$. The observation data $Y_k$ that we choose is $Y_k\mid \{V_{s}\}_{s\in[0,k]},\theta \sim \mathcal{N}(V_{k},\tau^2)$ (Gaussian distribution mean $V_{k}$ and variance $\tau^2$) where $\tau^2=0.01$.

\begin{figure}[H]
    \centering
    \includegraphics[width=0.9\textwidth, height=4.5cm]{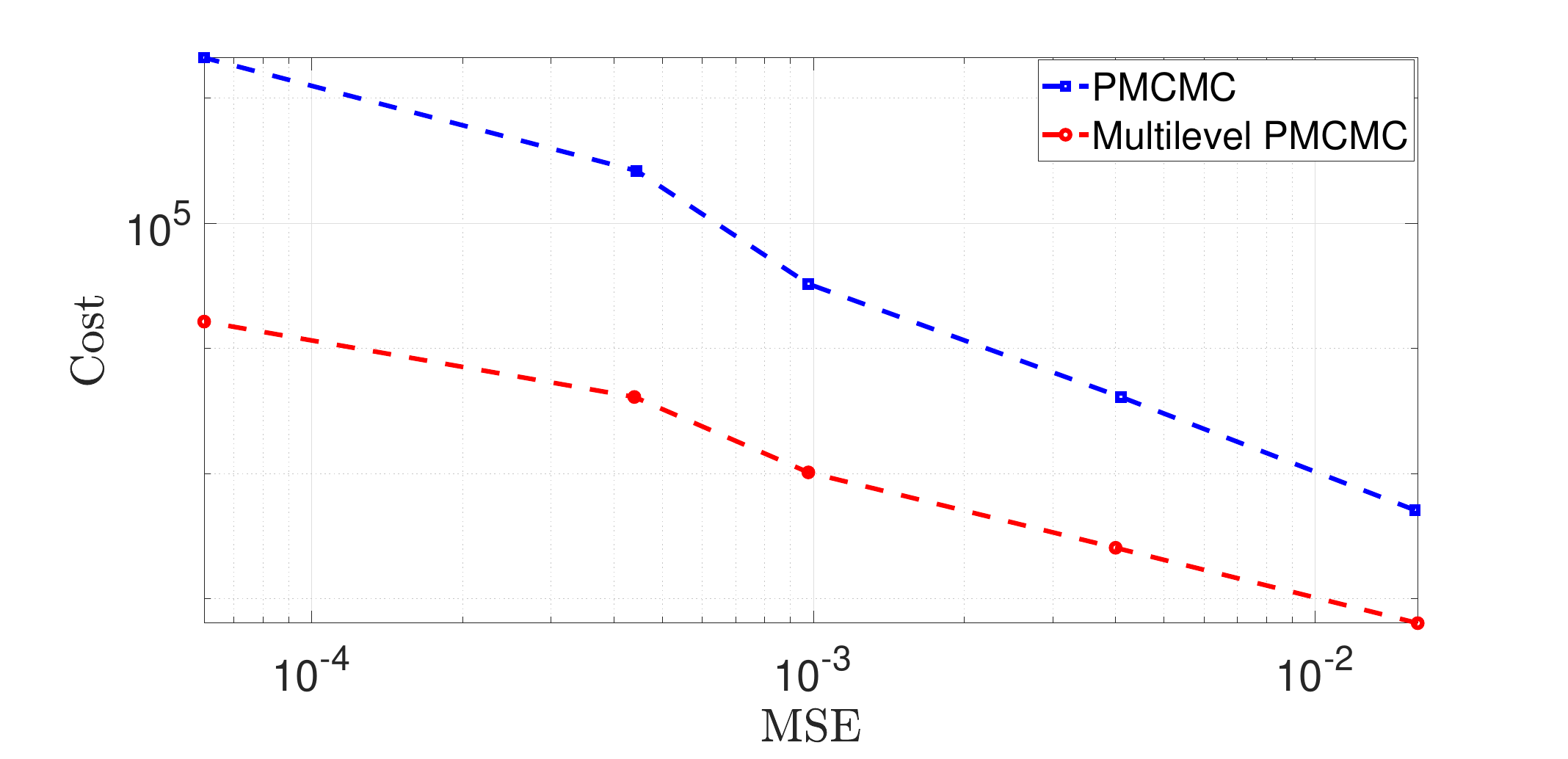}
    \caption{Mean-squared error versus cost function for the parameter of interest $S^{\text{dr}}$ in equation (\ref{moneqIF}).}
    \label{fig:MSEvsCostIF}
\end{figure}

For the MLPMCMC and PMCMC implementations, we choose $l\in \{3,4,5,6,7\}$,  and the iterations $M_l$ are chosen as above. The cost formulae for the MLPMCMC and PMCMC are:
\begin{align}
\label{eq:cost_mlmc}
\mathrm{Cost_{MLPMCMC}} &= \mathcal{O}\bigg(\sum^L_{l=3} M_l\Delta_l^{-1}\bigg), \\
\label{eq:cost_mc}
\mathrm{Cost_{PMCMC}} &= \mathcal{O}\bigg( M_0\Delta_L^{-1}\bigg).
\end{align}
 The true parameter $(S^{\text{dr}})^{\dagger}$ is generated from a high-resolution simulation with $L = 10$. The number of simulations (i.e.~repeats) we use for the MSE is $50$ and $T=100$. We set the number of particles in the PMCMC kernel to be $\mathcal{O}(T)$. We consistently used a fixed burn-in period of $1000$ iterations in all our simulations. The primary results of the mean-squared error-versus cost analysis are visually presented in Figure \ref{fig:MSEvsCostIF}.

\begin{figure}[H]
    \centering
    \subfloat[Autocorrelation function (ACF) of PMCMC chains for $S^{\text{dr}}$]{
        \includegraphics[width=0.9\textwidth, height=5cm]{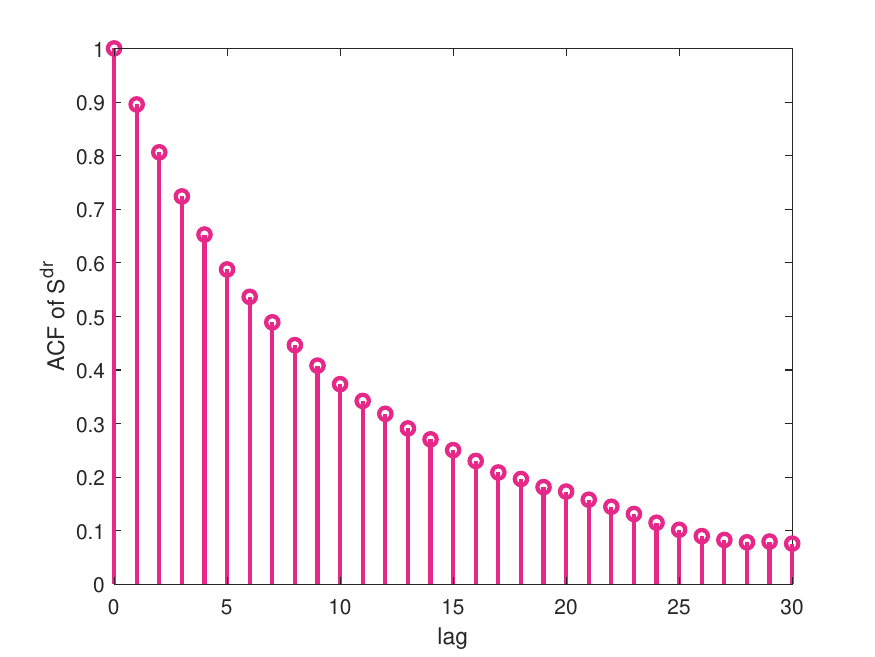}
        \label{fig:acf}
    } \vspace{0.3cm} \\

    \subfloat[Trace plot of estimated $S^{\text{dr}}$ with average indicated by the black line]{
        \includegraphics[width=0.9\textwidth, height=5cm]{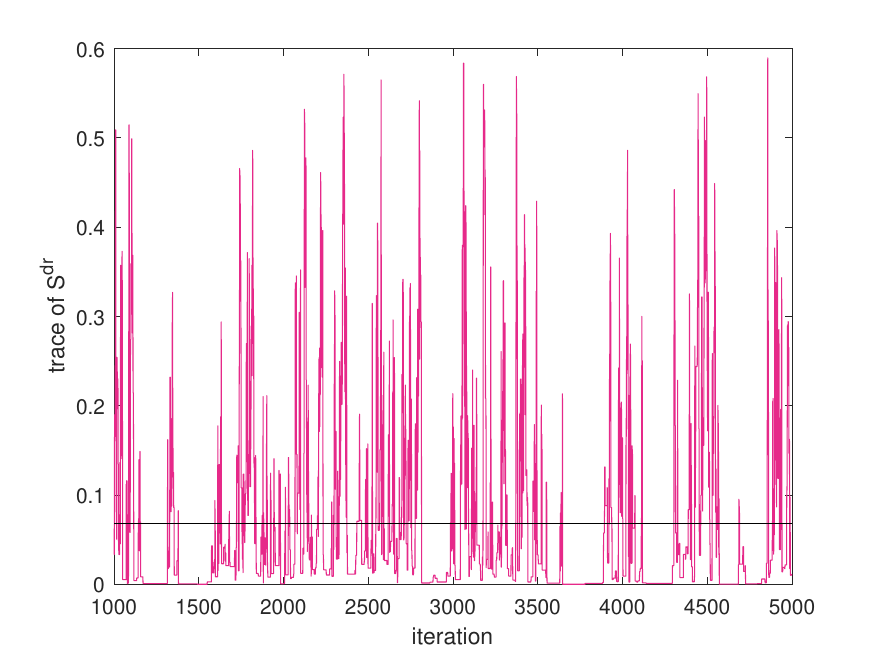}
        \label{fig:trace}
    } \vspace{0.3cm} \\

    \subfloat[Histogram of posterior distribution of the estimated parameter in (\ref{moneqIF})]{
        \includegraphics[width=0.9\textwidth, height=5cm]{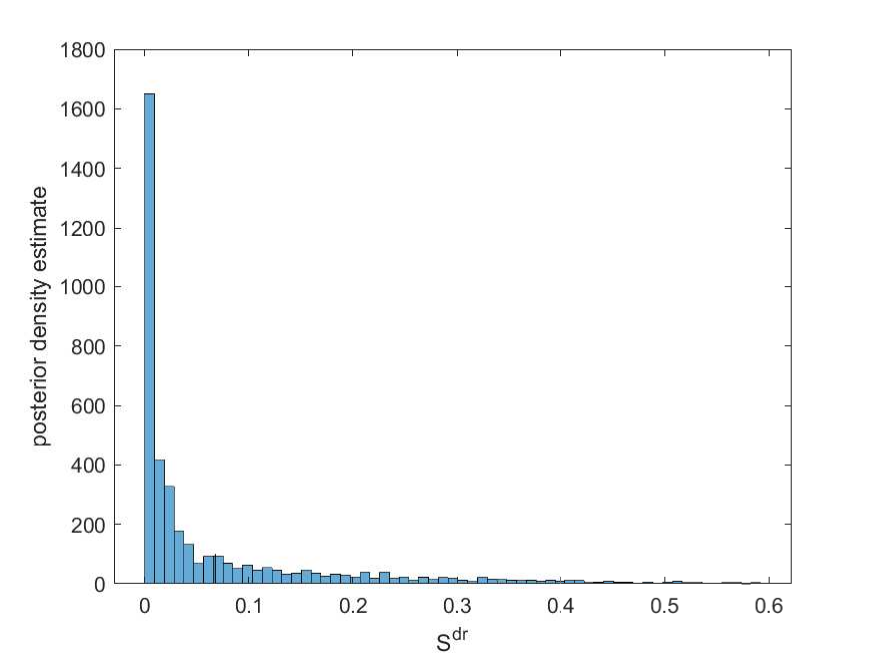}
        \label{fig:posterior}
    }
    
    \caption{PMMH outputs for our first model. (a) ACF plot for $S^{\text{dr}}$; (b) Trace plot for estimated $S^{\text{dr}}$, with the average value indicated by a black line; (c) Posterior distribution histogram of the estimated parameter of equation (\ref{moneqIF}).}
    \label{fig:Sdr}
\end{figure}


\subsubsection{Case 2: Small Network of Coupled \textcolor{red}{E}-\textcolor{blue}{I} Neurons}\label{sec:coupled_neurons}

We consider a small network of two neurons that are connected bidirectionally between excitatory and inhibitory cells. The system can be described by the following equations:
\begin{equation}
\left\{\begin{array}{lllclcll}
 \dif V_{1,t}=  \Big(  \frac{1}{\tau_{V}} ( V_{\text{reset}} - V_{1,t})  -  S^{\text{EI}}\, \sum_{k^{'}=1}^{n_t^{\text{I}}} \mathbb{I}_{[0,\infty)}(t - t_{k^{'}}^{\text{syn,I}})  \Big) \dif t + S^{\text{dr}} \dif N_{1,t} ,\vspace{1mm}\\
\dif V_{2,t}=  \Big(  \frac{1}{\tau_{V}} ( V_{\text{reset}} - V_{2,t})  +  S^{\text{IE}} \, \sum_{k^{'}=1}^{n_t^{\text{E}}} \mathbb{I}_{[0,\infty)}(t - t_{k^{'}}^{\text{syn,E}})  \Big) \dif t + S^{\text{dr}} \dif N_{2,t} ,\vspace{1mm}\\
\end{array}\right.
\label{moneqEI}
\end{equation}
where $\left\{ N_{1,t}\right\}_{t \geq 0}$ and $\left\{ N_{2,t}\right\}_{t \geq 0}$ are two one-dimensional homogeneous independent Poisson processes both with rate $\lambda = 0.8$.
In our case, we numerically simulate the model (\ref{moneqEI}) using a hybrid system formalism which exhibits discontinuities at firing times $\left\{t_{k^{'}}^{\text{syn,E}}\right\}_{k^{'}}^{}$ (number up-to time $t$ is $n_t^{\text{E}}$) and $\left\{t_{k^{'}}^{\text{syn,I}}\right\}_{k^{'}}^{}$ (number up-to time $t$ is $n_t^{\text{I}}$). One can define the times mathematically as follows:
\begin{eqnarray}
t_1^{\text{syn,Q}} & = & \inf\left\{t>0:V_{j(\text{Q}),t}\geq V_{\text{Thr}}\right\} \\
t_k^{\text{syn,Q}} & = & 
\inf\left\{t>t_{k-1}^{\text{syn,Q}}:V_{j(\text{Q}),t}\geq V_{\text{Thr}}\right\} \quad k\geq 2
\end{eqnarray}
where $\text{Q}\in\{\text{E,I}\}$ and $j(\text{I})=2$, $j(\text{E})=1$. Given this definition, we are indeed in the class of models given in Remark \ref{rem:path}.


The observations are such that for $k \in \left\{ 1,...,T \right\}$, 
$Y_k\mid \{V_s\}_{s\in[0,k]}, \theta \sim \mathcal{N}(V_k,\tau^2)$  and we set $\tau^2=0.02$. The parameters to be estimated are then $ \theta = (S^{\text{EI}}, \; S^{\text{IE}})^{\top} \in \mathbb{R}_{+}^{2}$ and these are both assigned independent Gamma priors that are $\mathcal{G}(0.005,0.005)$ for both parameters. The data is generated from a true parameter $\theta = (0.003,0.003)$. The number of samples utilized in the particle filter for the simulations is taken as $\mathcal{O}(T)$. In our multilevel approach, as for the first model, the only levels we choose are $l \in \left\{ 3,4,...,7 \right\}$.

\begin{figure}[H]
    \centering
    \subfloat[Trace plot of PMCMC chains for \( S^{IE} \)]{
        \includegraphics[width=0.9\textwidth, height=5cm]{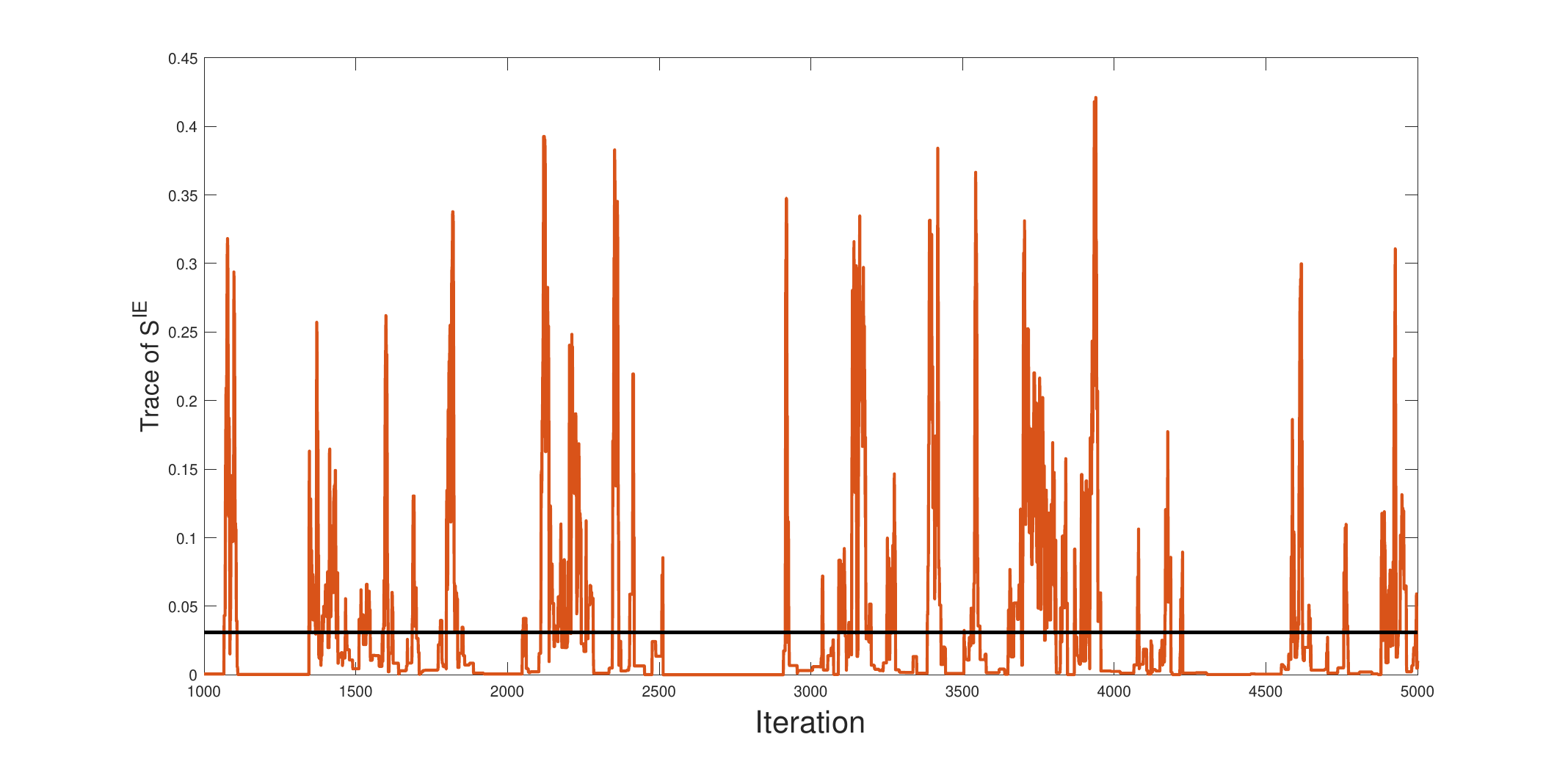}
        \label{fig:IterationSIE}
    } \\
    \subfloat[Posterior function plot for \( S^{IE} \)]{
        \includegraphics[width=0.9\textwidth, height=5cm]{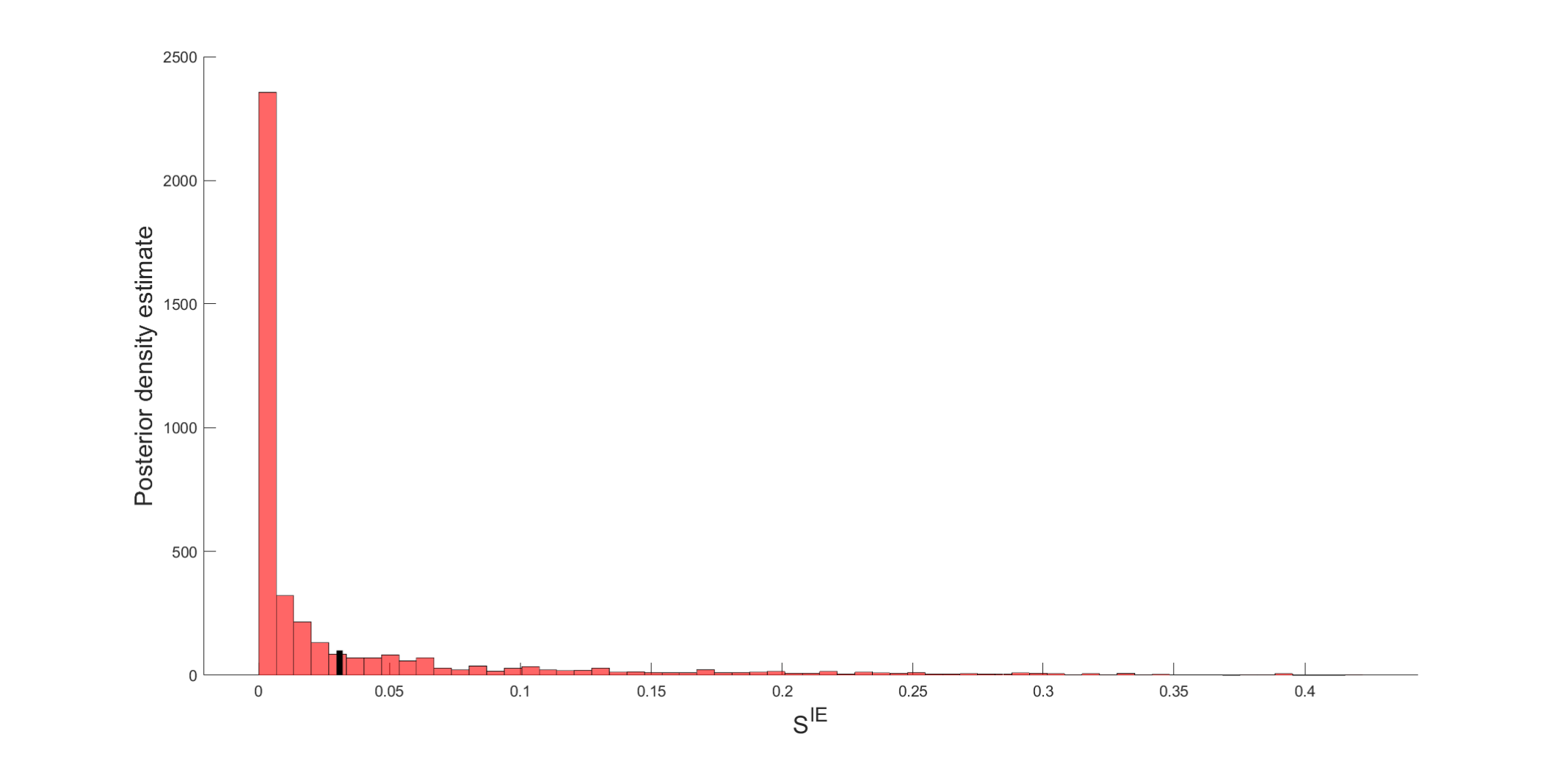}
        \label{fig:posteriorSIE}
    } \\
    \subfloat[Mean-squared error vs. Cost function for \( S^{IE} \)]{
        \includegraphics[width=0.9\textwidth, height=5cm]{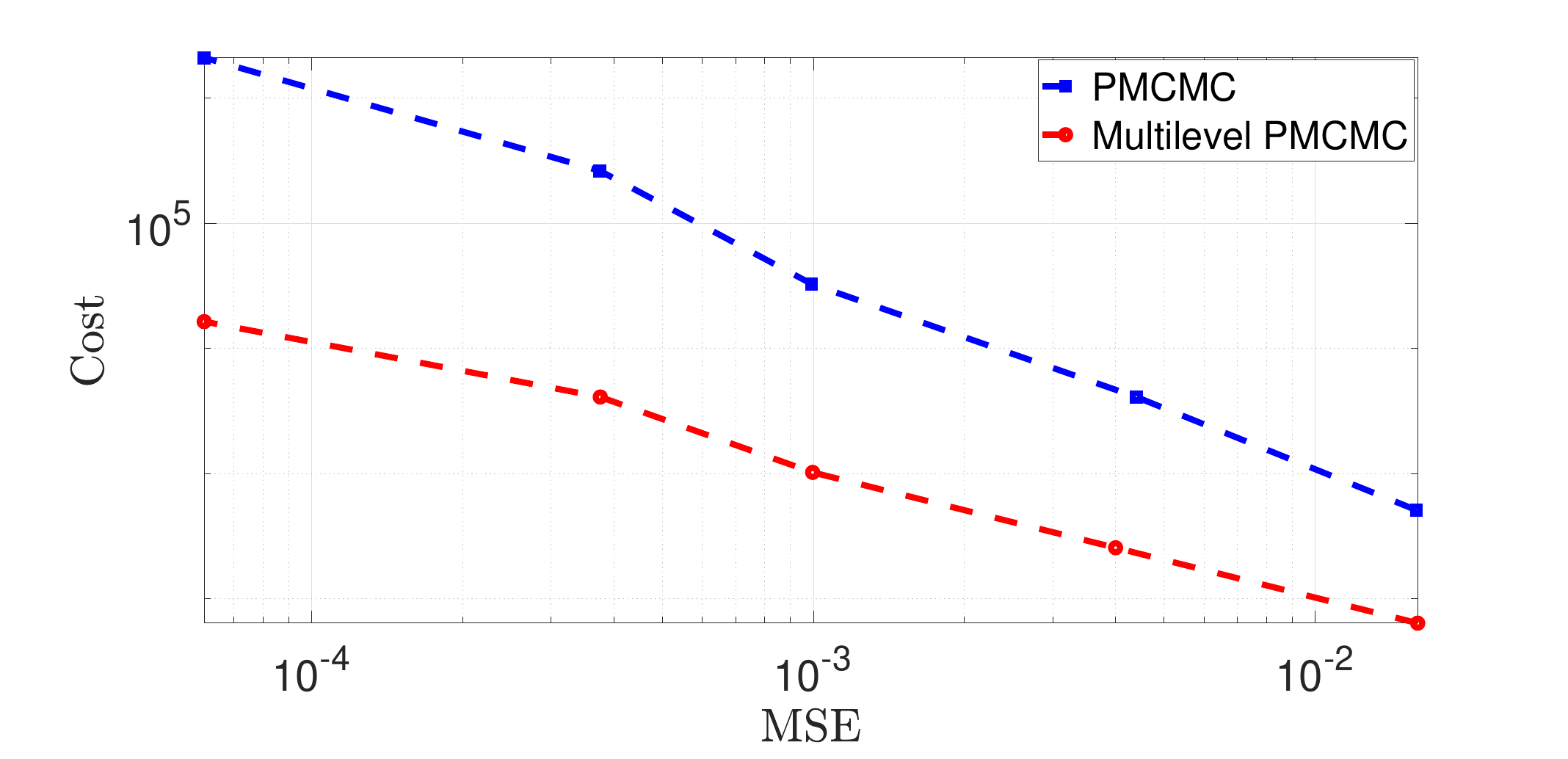}
        \label{fig:msevscostSIE}
    }
    \caption{Simulations output of equation (\ref{moneqEI}). (a) Trace plot of PMCMC chains for \( S^{IE} \); (b) Posterior function plot; (c) MSE vs. Cost plot.}
    \label{fig:SIE}
\end{figure}

\begin{figure}[H]
    \centering
    \subfloat[(a) Trace plot of PMCMC chains for \( S^{IE} \)]{
        \includegraphics[width=0.9\textwidth, height=5cm]{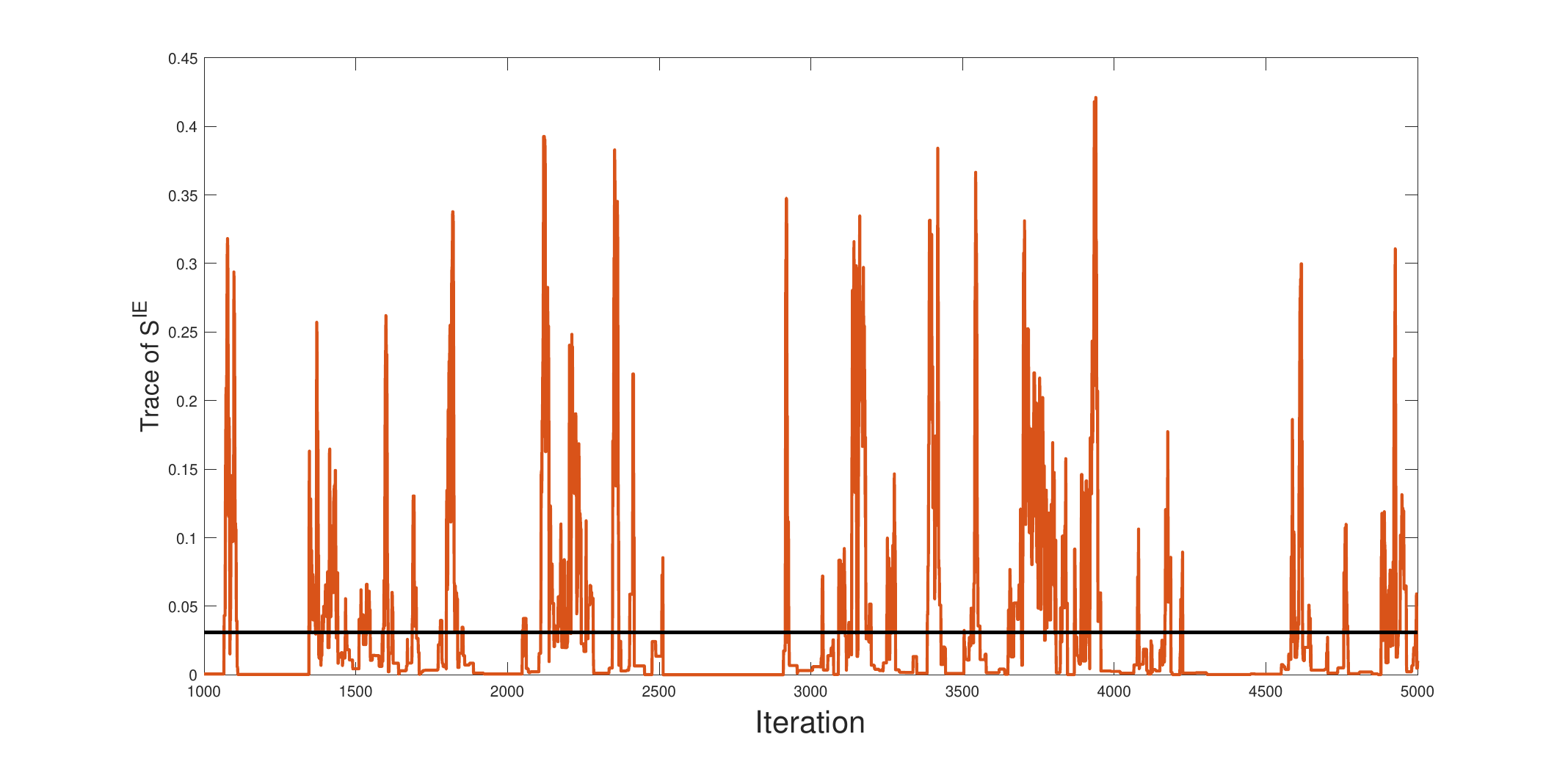}
        \label{fig:IterationSIE}
    } \vspace{0.3cm} \\
    
    \subfloat[(b) Posterior function plot for \( S^{IE} \)]{
        \includegraphics[width=0.9\textwidth, height=5cm]{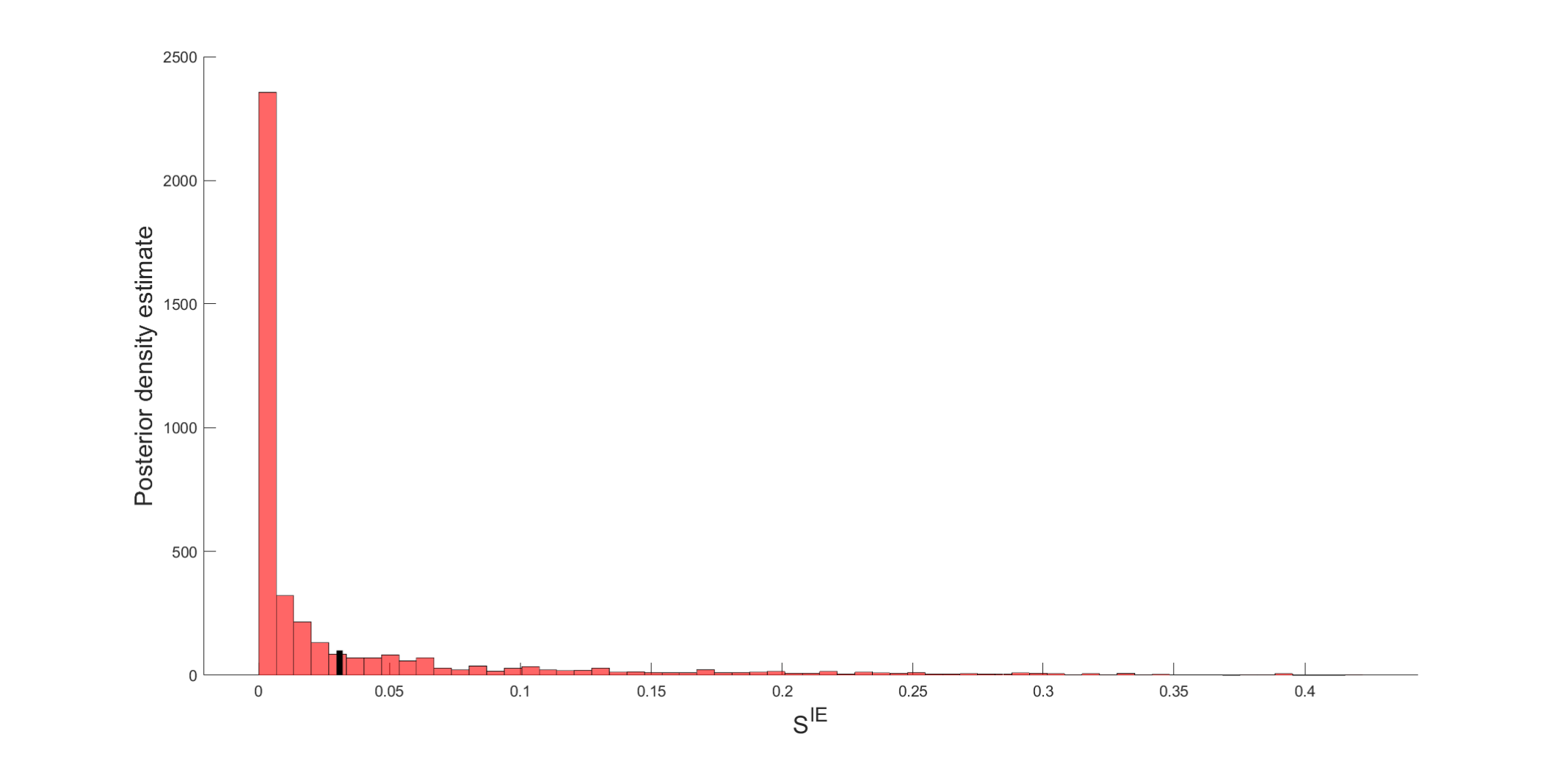}
        \label{fig:posteriorSIE}
    } \vspace{0.3cm} \\
    
    \subfloat[(c) Mean-squared error vs. Cost function for \( S^{IE} \)]{
        \includegraphics[width=0.9\textwidth, height=5cm]{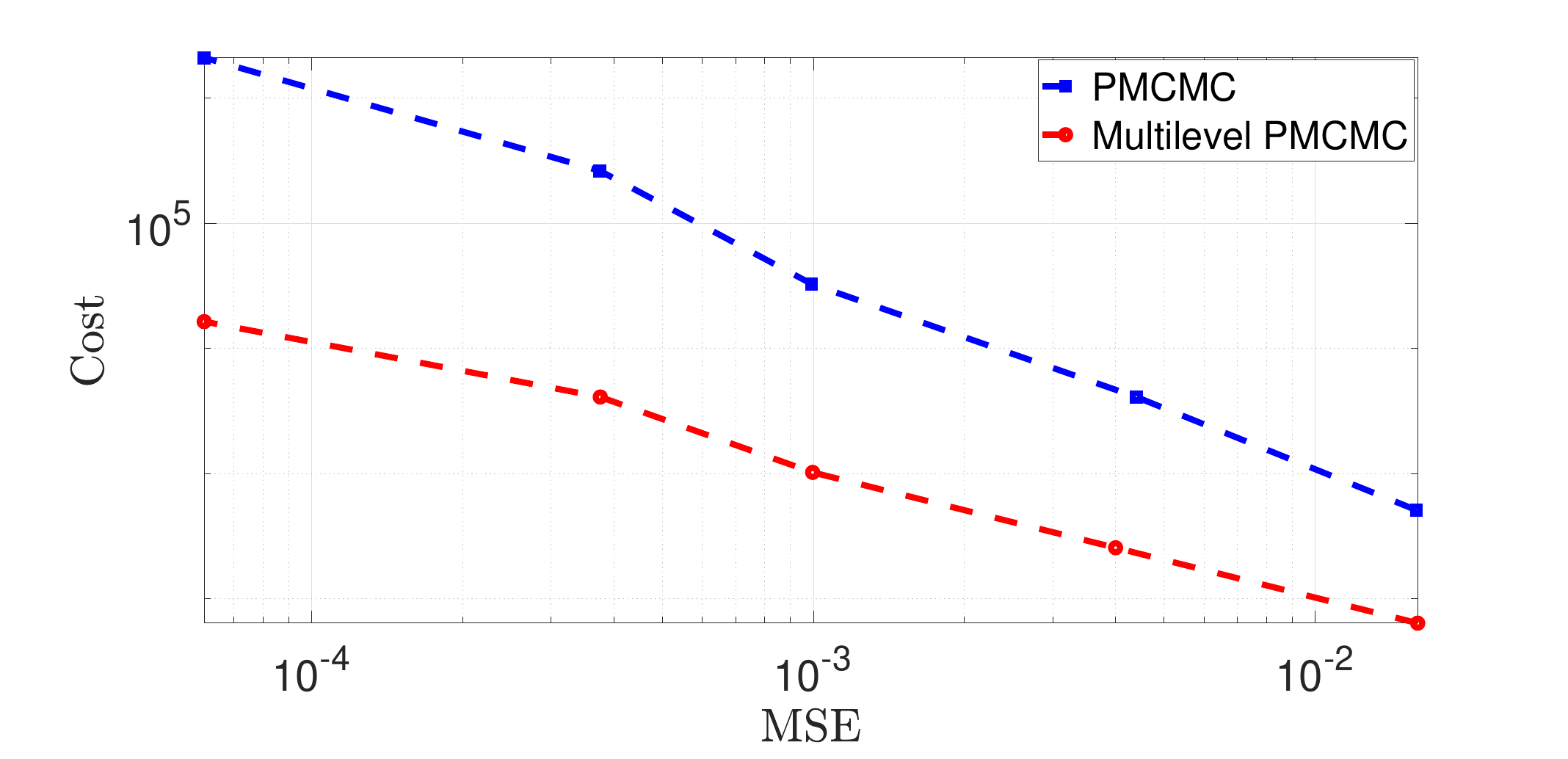}
        \label{fig:msevscostSIE}
    }
    
    \caption{Simulation outputs of equation (\ref{moneqEI}): (a) Trace plot of PMCMC chains for \( S^{IE} \); (b) Posterior function plot; (c) Mean-squared error vs. Cost function plot.}
    \label{fig:SEI}
\end{figure}

We first provide numerical results of the parameter of interest, $S^{\text{IE}}$. In Figure \ref{fig:SIE} we can observe some output from the single-level MCMC algorithm when it was executed at level $7$. We observe both the histograms and trace plots generated by the Markov chain and also we can see based upon 50 repeats, the cost versus MSE plots. It shows that the multilevel Monte Carlo MCMC method incurs a lower cost for achieving a mean squared error. In Table \ref{tab:S} we estimate the rates, that is, log cost against log MSE based upon Figure \ref{fig:SIE}. This implies that a single-level algorithm incurs a cost of $\mathcal{O}(\epsilon^{-3})$ and a multilevel has a cost of $\mathcal{O}(\epsilon^{-2})$  to achieve an MSE of $\mathcal{O}(\epsilon^{2})$.
We repeat the experiments now for the second parameter of interest $S^{\text{EI}}$. Figure \ref{fig:SEI} and Table \ref{tab:S} show the performance of the multilevel MCMC method (again single-level MCMC at level 7) and in terms of performance is indeed in line with our expectations.

\begin{table}[H]
\begin{center}
\begin{tabular}{ c c c c } 
\hline \hline
Model  & Parameter & PMCMC & MLPMCMC  \\
\hline
\hline
\textbf{Case 1: Driving Integrate-and-Fire} & $S^{\text{dr}}$ &-1.54 &-1.02  \\ 

 \hline

\multirow{2}{*}{\textbf{Case 2: Small Network of E-I Coupled Neurons}} & $S^{\text{EI}}$   & -1.47  &  -1.05 \\

  &  $S^{\text{IE}}$   & -1.52 & -1.02 \\

 \hline
 \hline
\end{tabular}
\caption{Estimated rates of convergence of mean-squared error with respect to the cost function for the three key parameters $ \big( S^{\text{dr}}, \; S^{\text{EI}}, \; S^{\text{IE}} \big) $, adapted to the curves simulated above. MLPMCMC is a multilevel PMCMC.}
\label{tab:S}
\end{center}
\end{table}

\subsection{\textcolor{black}{ Inference of the Izhikevich parameters from real action potential data}}

In this section we estimate the parameters of interest of the stochastic Izhikevich model using real membrane voltage data recorded from rats.

\subsubsection{Biological Background of the Izhikevich Model}

The Izhikevich model \cite{Izhi} is widely used in computational neuroscience due to its ability to accurately replicate the spiking behavior of neurons with relatively simple mathematical formulations. The model is designed to capture essential features of neuronal firings, such as regular spiking, bursting, and fast-spiking while maintaining computational efficiency. The model consists of two main variables:
\begin{itemize}
    \item \(v\): the membrane potential of the neuron.
    \item \(u\): a recovery variable representing the combined effect of ion channel dynamics.
\end{itemize}

The deterministic equations governing the neuron dynamics are given by:
\[
\frac{dv}{dt} = 0.04v^2 + 5v + 140 - u + I(t),
\]

\[
\frac{du}{dt} = a(bv - u),
\]

where \(I(t)\) is the external input current, and the parameters \(a\), \(b\), \(c\), and \(d\) define the specific type of neuronal behavior (e.g., regular spiking, fast spiking). When the membrane potential \(v\) reaches a certain threshold (typically 30 mV), a reset occurs:
\begin{equation}
    v \leftarrow c, \quad u \leftarrow u + d,
\end{equation}
representing the neuron firing an action potential (spike). This simple yet versatile model can reproduce a wide range of firing patterns seen in biological neurons, making it a popular choice in studies of neuronal networks and dynamical systems.

\subsubsection{Stochastic Izhikevich Model with Poisson-driven Inputs}

To simulate the impact of random synaptic inputs or external stimuli on a neuron's behavior, we incorporate a Poisson-driven stochastic component into the deterministic Izhikevich model. In this framework, random input events, such as synaptic spikes from other neurons, are modeled using a homogeneous Poisson process with rate \(\lambda > 0\). The modified Izhikevich model, incorporating these stochastic inputs, is expressed as:

\begin{equation}
    \dif v_t = \left(0.04v_t^2 + 5v_t + 140 - u_t + I_{ext}\right)\dif t + S^{\text{dr}} \dif N_t,
\end{equation}
where \(S^{\text{dr}}\) is a scaling factor controlling the strength of the Poisson-driven input. This term introduces random discrete jumps in the membrane potential \(v_t\), reflecting the stochastic nature of synaptic events in biological neurons. The recovery variable \(u_t\) follows the original deterministic equation:

\begin{equation}
    \dif u_t = a(bv_t - u_t)\dif t.
\end{equation}

Thus, the complete system of equations becomes:
\begin{equation}
\begin{aligned}
    \dif v_t &= \left(0.04v_t^2 + 5v_t + 140 - u_t + I(t)\right)\dif t + S^{\text{dr}} \dif N_t, \\
    \dif u_t &= a(bv_t - u_t)\dif t,
\end{aligned}
\label{eqP}
\end{equation}
with the Poisson process introducing random fluctuations in the membrane potential, while the recovery variable evolves according to the original deterministic dynamics. The role of stochasticity in shaping neuronal firing behavior is explored by varying the strength of the Poisson-driven noise, \(S^{\text{dr}}\). This factor modulates the intensity of random spikes, influencing both the membrane potential and the power spectral density (PSD) of the neuron's firing. We simulate three different values of \(S^{\text{dr}}\): 0.1, 0.5, and 1, which represent low, moderate, and high levels of noise, respectively.

\begin{figure}[H]
    \centering
    \includegraphics[width=0.8\textwidth]{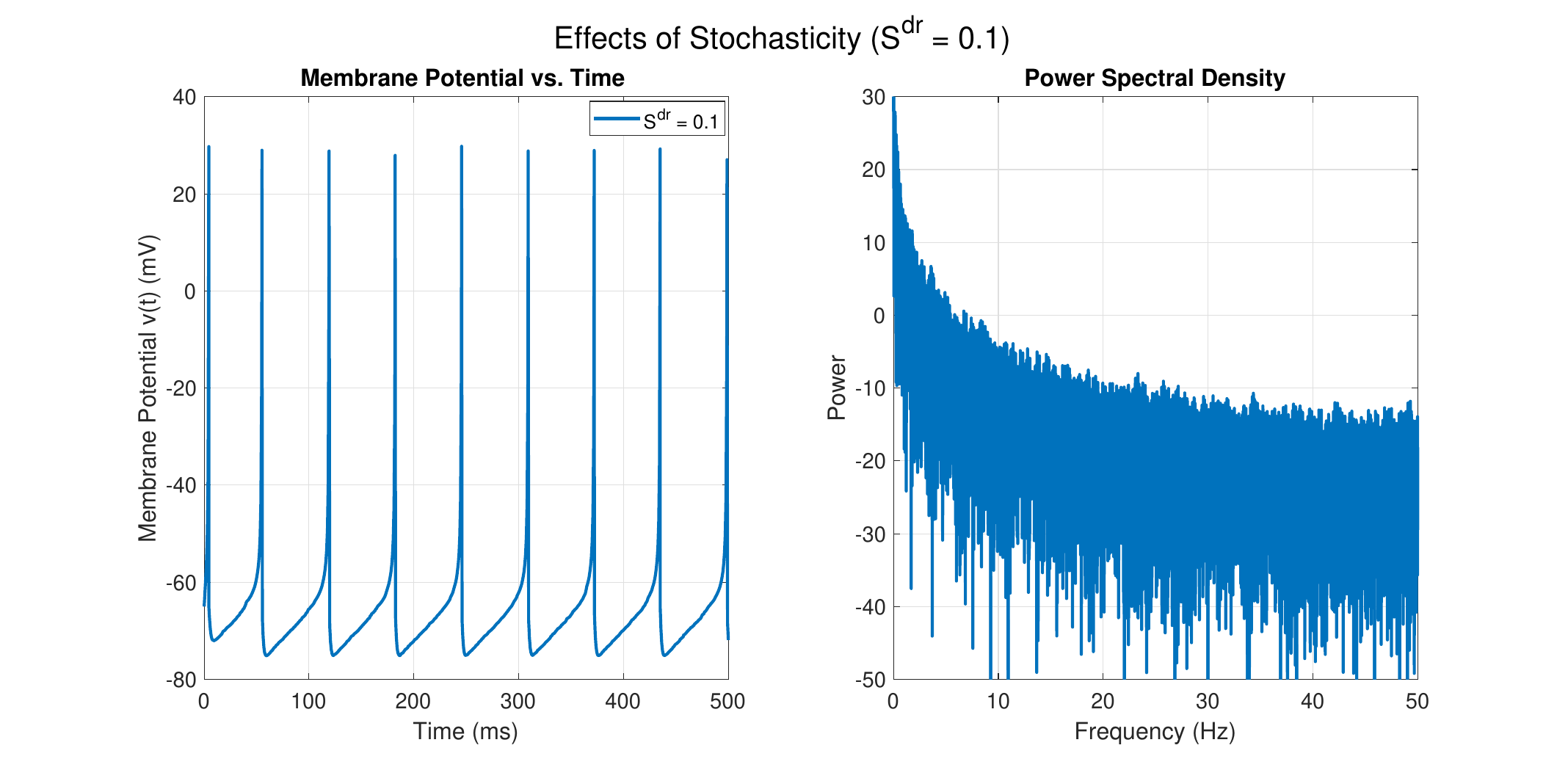}
    \caption{Membrane Potential vs. Time and Power Spectral Density for \( S^{dr} = 0.1 \)}
    \label{fig:Sdr_0.1}
\end{figure}

\begin{figure}[H]
    \centering
    \includegraphics[width=0.8\textwidth]{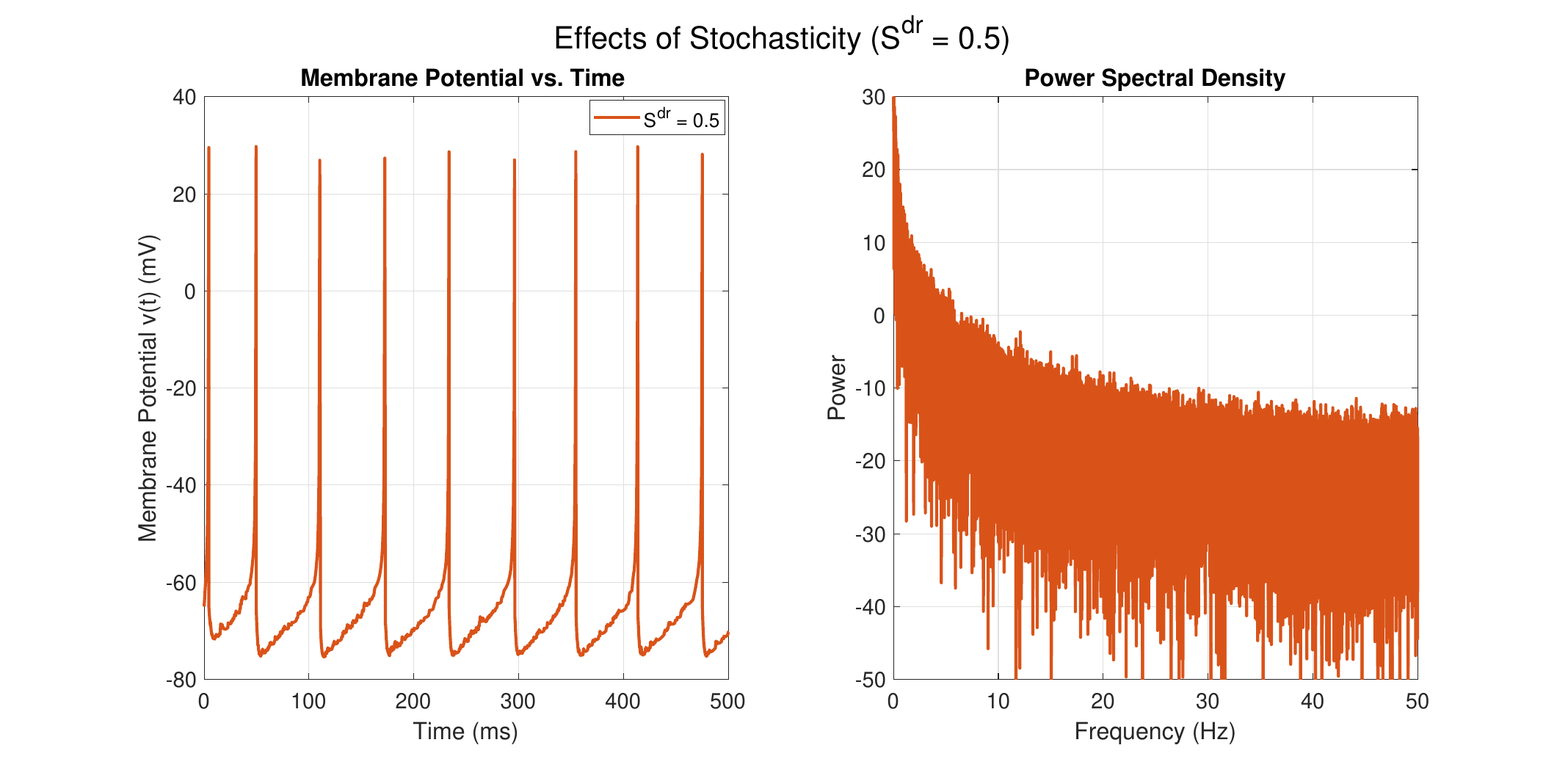}
    \caption{Membrane Potential vs. Time and Power Spectral Density for \( S^{dr} = 0.5 \)}
    \label{fig:Sdr_0.5}
\end{figure}

\begin{figure}[H]
    \centering
    \includegraphics[width=0.8\textwidth]{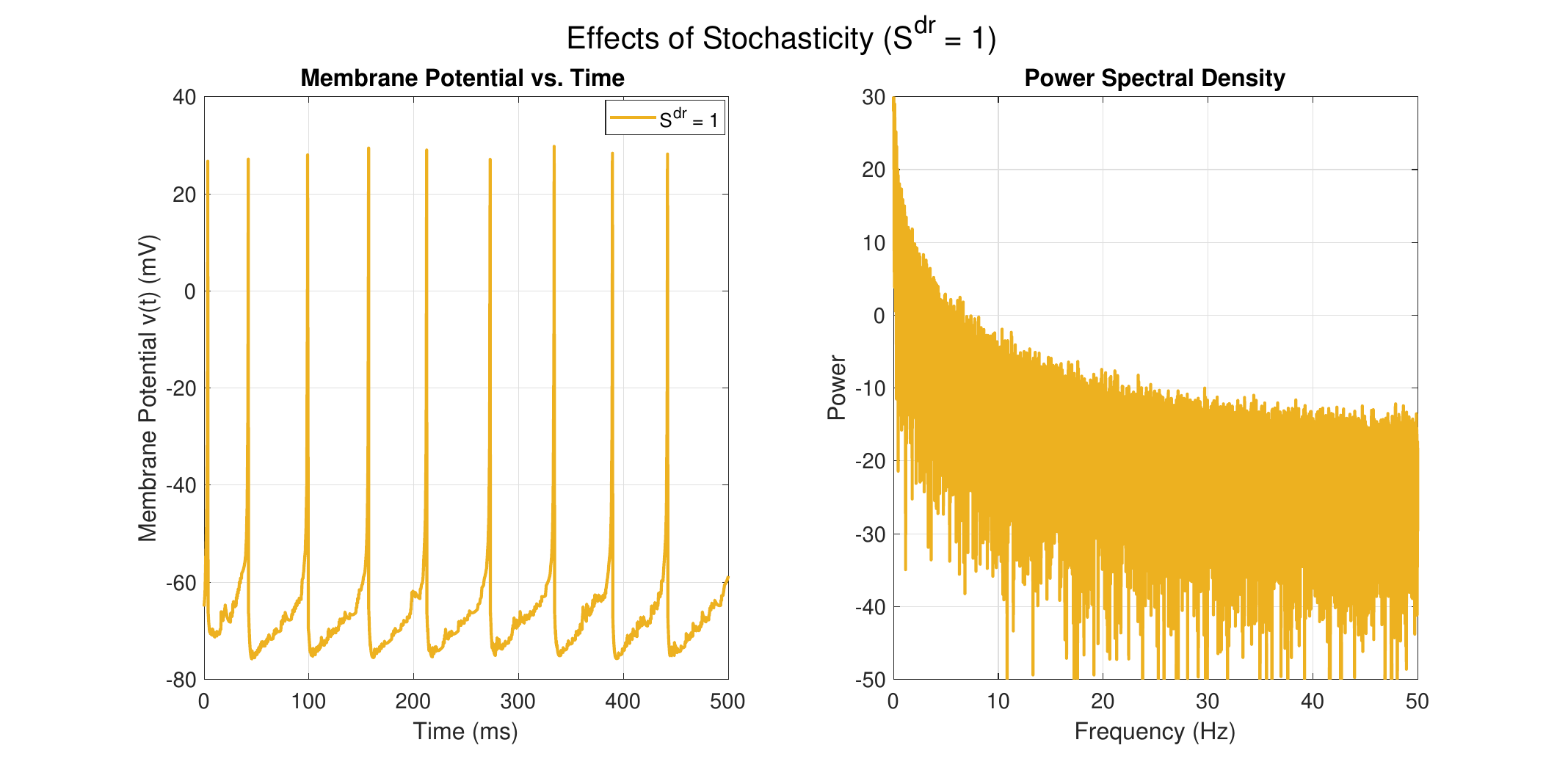}
    \caption{Membrane Potential vs. Time and Power Spectral Density for \( S^{dr} = 1 \)}
    \label{fig:Sdr_1}
\end{figure}

Figures \ref{fig:Sdr_0.1} - \ref{fig:Sdr_1} illustrate the membrane potential over time and the corresponding PSDs for these three values of \(S^{\text{dr}}\). At lower levels of stochasticity (e.g., \(S^{\text{dr}} = 0.1\)), the neuron exhibits regular spiking behavior, resembling periodic firing patterns. As \(S^{\text{dr}}\) increases to 0.5 and 1, the neuron shows more irregular and unpredictable firing, with spiking events occurring more randomly. This shift in firing behavior reflects the increased influence of stochastic input, which disrupts the regular oscillations of the membrane potential.

The power spectral density analysis further emphasizes this transition. As the strength of stochasticity rises, the frequency spectrum broadens, indicating a greater range of oscillatory activity. This broadening of the frequency spectrum is a signature of increased variability in the neuron's firing pattern, highlighting the neuron's adaptation to more chaotic input conditions. We observe a clear shift in the power distribution as \(S^{\text{dr}}\) increases. From a neuroscience perspective, this stochastic behavior mimics the random fluctuations inherent in biological neural systems, such as synaptic noise, intrinsic neuronal noise, or the irregular nature of sensory inputs. These random inputs can enhance the neuron's ability to respond flexibly to dynamic environments, playing a crucial role in processes such as sensory processing, attention, and decision-making. The parameters of the model-\(a = 0.02\) (recovery time constant), \(b = 0.2\) (recovery sensitivity), \(c = -65\) mV (membrane potential reset value), and \(d = 8\) (recovery variable reset value)-are critical for shaping the neuron's intrinsic firing properties. The parameter \(a\) governs the neuron's recovery dynamics, while \(b\) influences the sensitivity of the recovery variable to the membrane potential. The reset parameters \(c\) and \(d\) determine how the neuron resets after a spike, influencing the firing rate and regularity. These intrinsic parameters must be carefully tuned to ensure that the model captures realistic neuronal behavior under varying levels of noise. This combination of deterministic and stochastic dynamics provides a more comprehensive representation of neuronal activity, reflecting the importance of both intrinsic properties and external noise in shaping brain function. The figures presented provide a visual representation of how the interplay between these factors leads to changes in firing patterns and spectral properties, further highlighting the role of stochasticity in neuronal behavior.

\subsubsection{Overview of Collected Data}

The dataset used in this study comprises neural recordings from the fifth lumbar (L5) dorsal rootlet of a single adult female Sprague Dawley rat, obtained through a neural recording experiment using a custom-made electrode array. The dataset is accessible via \href{https://data.mendeley.com/datasets/ybhwtngzmm/1}{Mendeley Data}, with full experimental details described in \cite{DataRat}. It includes 20 recordings, each lasting $250 \; ms$ and sampled at $50 \; kHz$, providing high-resolution voltage data across five channels. Recordings were conducted under two conditions: resting (10 recordings) and cutaneous stimulation (10 recordings), with stimulation targeting the dermatome to activate associated sensory axons in the dorsal root. 
In this paper, we selected four representative recordings-1551 and 1608 (resting) and 1554 and 1612 (stimulation)-and focused on Channel 1, the most distal electrode closest to the tail, which consistently provided a clear signal. This targeted selection enables a focused comparison of neural activity between resting and stimulated states while preserving experimental integrity.

\subsubsection{Parameter Inference using PMCMC}

We infer the parameters $\theta = (a, b, c, d, S^{dr}) \in \mathbb{R}^5$ of the stochastic Izhikevich model (\ref{eqP}) from real data consisting of membrane voltage recordings obtained from rats. The parameter inference is performed using a Particle Metropolis-Hastings algorithm, a variant of Particle Markov Chain Monte Carlo (PMCMC). The stochastic Izhikevich model is fitted to four representative datasets (1551, 1608, 1554, and 1612) with a time horizon $T = 250 \; ms$. The number of samples used in the particle filter for the simulations is of the order \( O(T) \). In the Particle Metropolis-Hastings algorithm, the step size is chosen as \( \Delta_l = 2^{-l} \) with \( l = 9 \). The algorithm runs for \( 10^4 \) iterations, with the first \( 10^3 \) iterations serving as a burn-in period to discard transient samples and ensure convergence. Adaptive resampling is applied using the systematic resampling method, which dynamically adjusts the particle set based on their weights, effectively preventing particle degeneracy. To evaluate the diversity of the particles, we calculate the effective sample size (ESS) after resampling and keep it below half the total number of particles. This helps the algorithm explore the parameter space well while avoiding overfitting, which is important for accurately modeling neural dynamics and the underlying randomness.

In this context, the choice of prior distributions plays a crucial role in guiding the algorithm's exploration of the parameter space. For the parameter \(a\), we choose a \textit{log-normal prior}, reflecting our belief that recovery time constants in biological neurons tend to follow a positively skewed distribution, with most values clustering around a central value but with potential for larger values in some neuron types. For the parameter \(b\), we select a \textit{uniform prior} within a biologically plausible range, which reflects the assumption that the recovery sensitivity is equally likely to take any value within that range, based on our lack of specific prior knowledge. For the membrane potential reset parameter \(c\) and the recovery variable reset parameter \(d\), we opt for \textit{normal priors}. The choice of normal priors reflects the fact that both parameters are expected to center around typical values observed in real neurons (e.g., around \(-65\) mV for \(c\) and 8 for \(d\)), with some variability. Finally, for the Poisson-driven input strength \(S^{\text{dr}}\), we select a \textit{Gamma distribution} to model the strength of random input, as it can effectively capture the range of values from small to large noise levels, appropriate for the stochastic processes that influence neuronal firing. Here, we use two sets of priors

\begin{table}[ht]
\centering
\normalsize 
\setlength{\tabcolsep}{12pt} 
\renewcommand{\arraystretch}{1.5} 
\begin{tabular}{||l||c||c||}
\hline
\textbf{Parameter} & \textbf{Resting (1551, 1608)} & \textbf{Stimulated (1554, 1612)} \\
\hline
\(a\) (Recovery time) & \(\log\mathcal{N}(-3, 0.5^2)\) & \(\log\mathcal{N}(-2.5, 0.6^2)\) \\
\hline
\(b\) (Recovery sensitivity) & \(\mathcal{U}(0.1, 0.5)\) & \(\mathcal{U}(0.2, 0.6)\) \\
\hline
\(c\) (Reset membrane potential) & \(\mathcal{N}(-65, 5^2)\) & \(\mathcal{N}(-60, 5^2)\) \\
\hline
\(d\) (Reset recovery variable) & \(\mathcal{N}(8, 1^2)\) & \(\mathcal{N}(9, 1^2)\) \\
\hline
\(S^{\text{dr}}\) (Poisson noise strength) & \(\Gamma(2, 0.5)\) & \(\Gamma(3, 0.7)\) \\
\hline
\end{tabular}
\caption{Summary of priors for parameter inference under resting and stimulated cases.}
\label{table:priors}
\end{table}

The role of these prior distributions in Table \ref{table:priors} is pivotal in the PMCMC algorithm, particularly during the acceptance-rejection step, where candidate parameter values are evaluated based on their likelihood given the observed data. As the PMCMC algorithm iterates, the prior distributions are used to update the weights of the particles at each iteration, adjusting the exploration of the parameter space. This ensures that the algorithm prioritizes more probable regions, based on both the likelihood of the data and the constraints imposed by the priors, thereby avoiding over-exploration of unrealistic parameter values and improving the efficiency of the parameter inference process.

Alternative prior choices could be considered. For example, one might use a Gamma distribution for \(a\), a log-uniform prior for \(b\), or exponential priors for \(c\) and \(d\). While such variations could alter the parameter space that the algorithm explores, their impact on the final parameter estimates is expected to be minimal, as long as the priors remain within reasonable biological bounds. The current priors, however, are both informed by biological knowledge and computationally efficient, providing a robust foundation for parameter inference that aligns with known properties of neuronal dynamics.

\begin{figure}[H]
    \centering
    \includegraphics[width=0.8\textwidth]{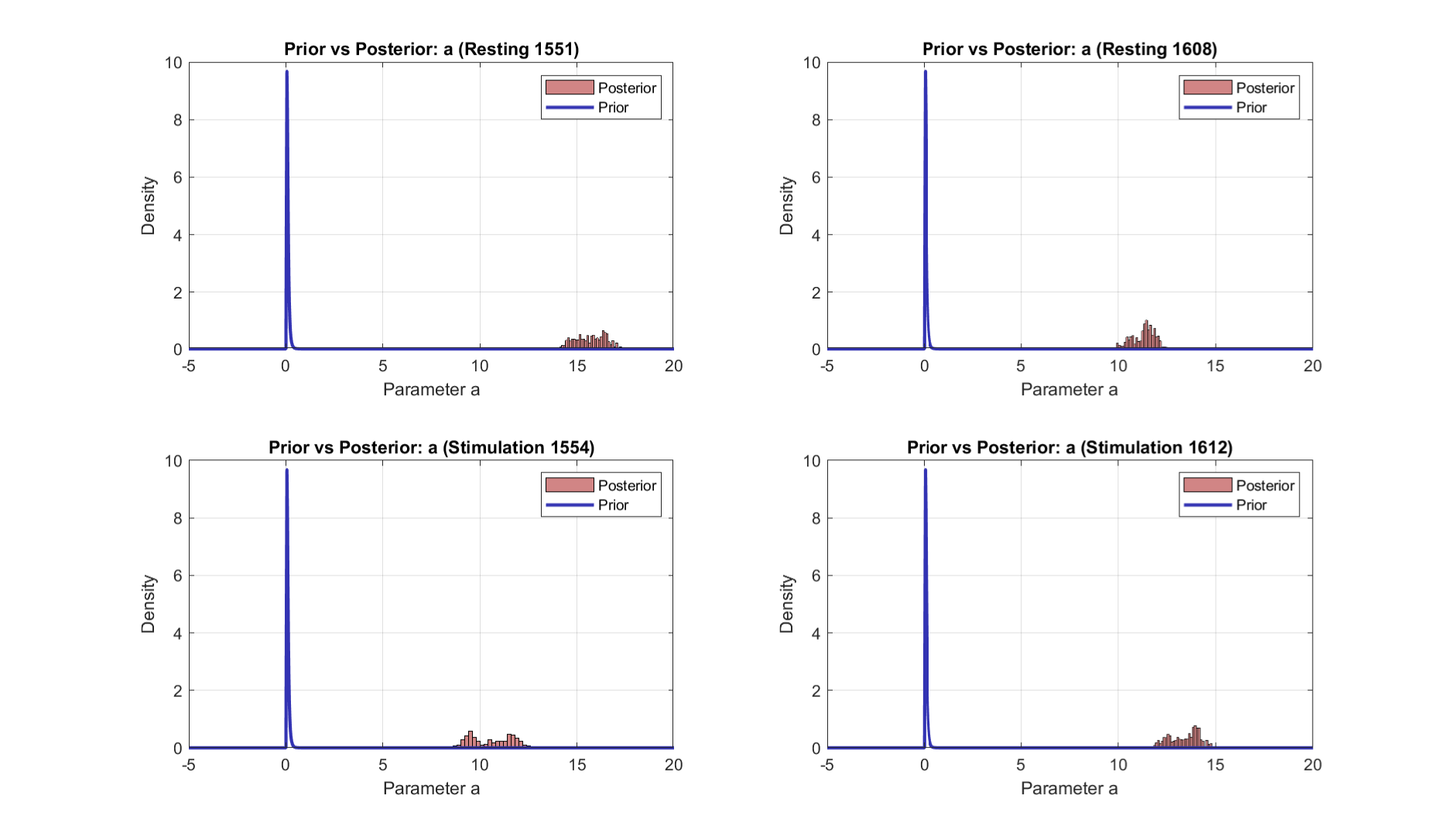}
    \caption{Posterior and Prior Distributions for Parameter \(a\) derived from Resting and Stimulated Real Data. The figure compares the posterior distributions (red) with the prior distributions (blue) for the parameter \(a\).}
    \label{fig:posteriora}
\end{figure}

\begin{figure}[H]
    \centering
    \includegraphics[width=0.8\textwidth]{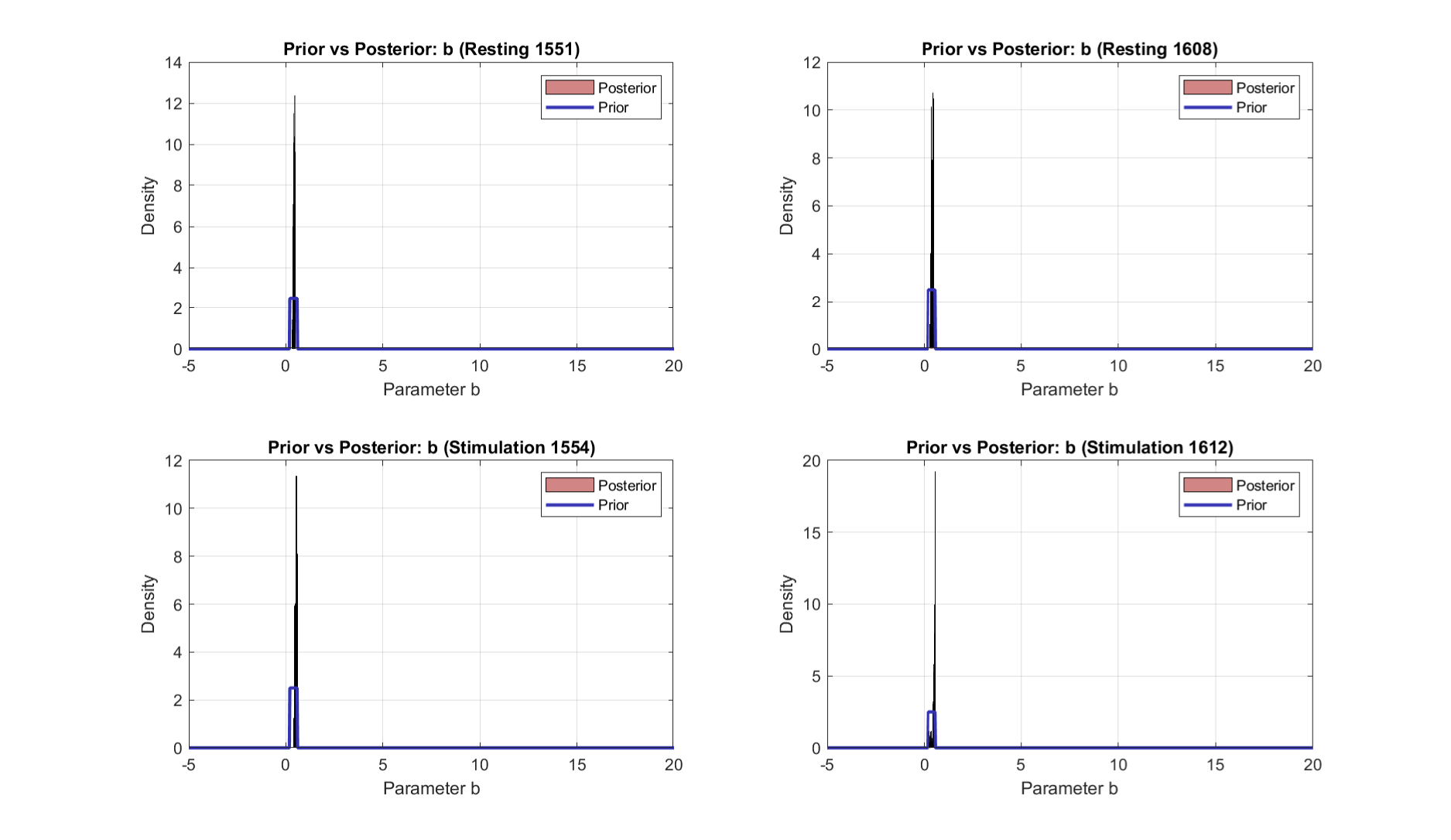}
    \caption{Posterior and Prior Distributions for Parameter \(b\) derived from Resting and Stimulated Real Data. The figure highlights the shift from the prior to the posterior for the parameter \(b\).}
    \label{fig:posteriorb}
\end{figure}

\begin{figure}[H]
    \centering
    \includegraphics[width=0.8\textwidth]{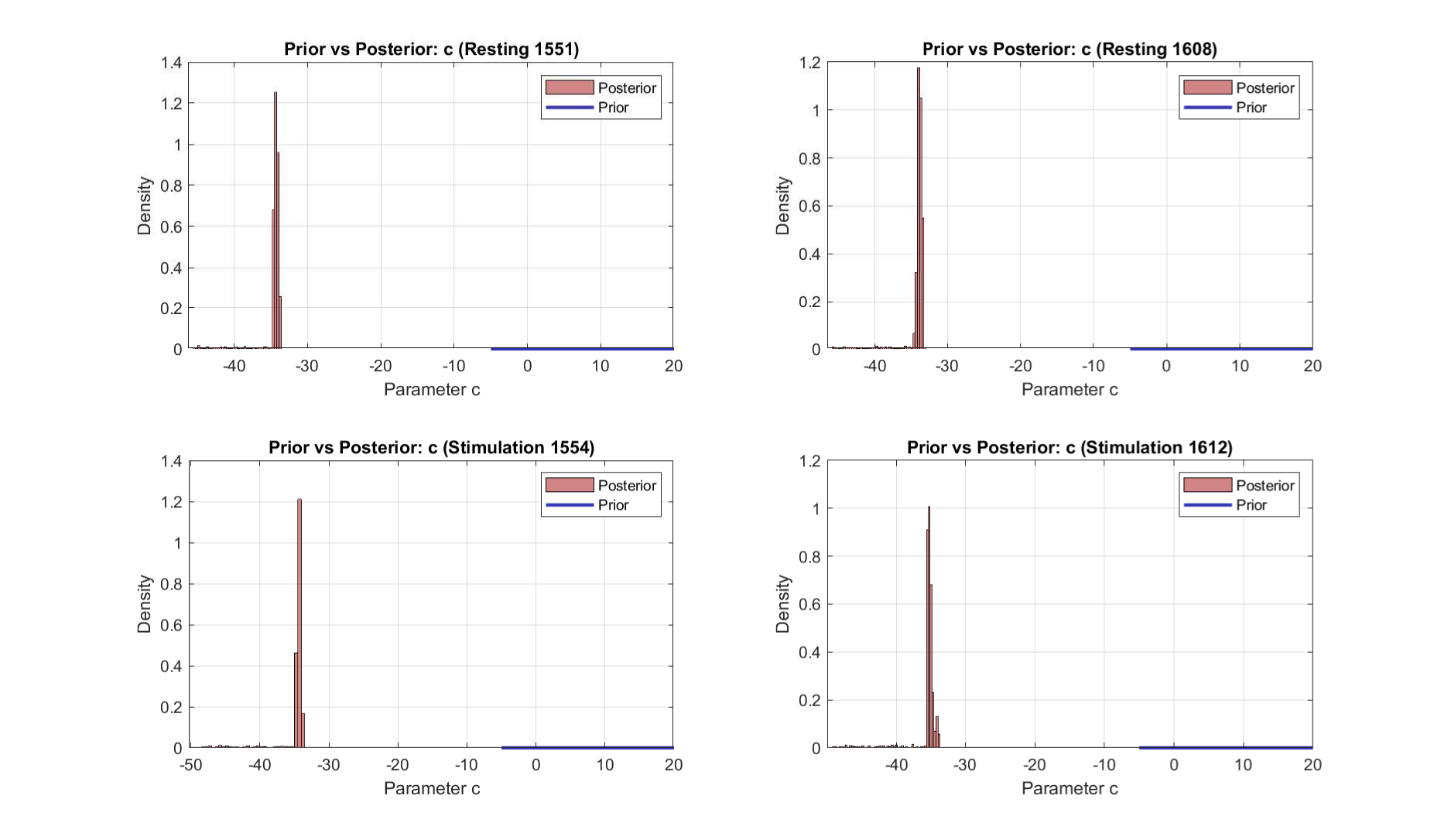}
    \caption{Posterior and Prior Distributions for Parameter \(c\) derived from Resting and Stimulated Real Data. The posterior distributions provide insights into the inferred reset membrane potential values for parameter \(c\).}
    \label{fig:posteriorc}
\end{figure}

\begin{figure}[H]
    \centering
    \includegraphics[width=0.8\textwidth]{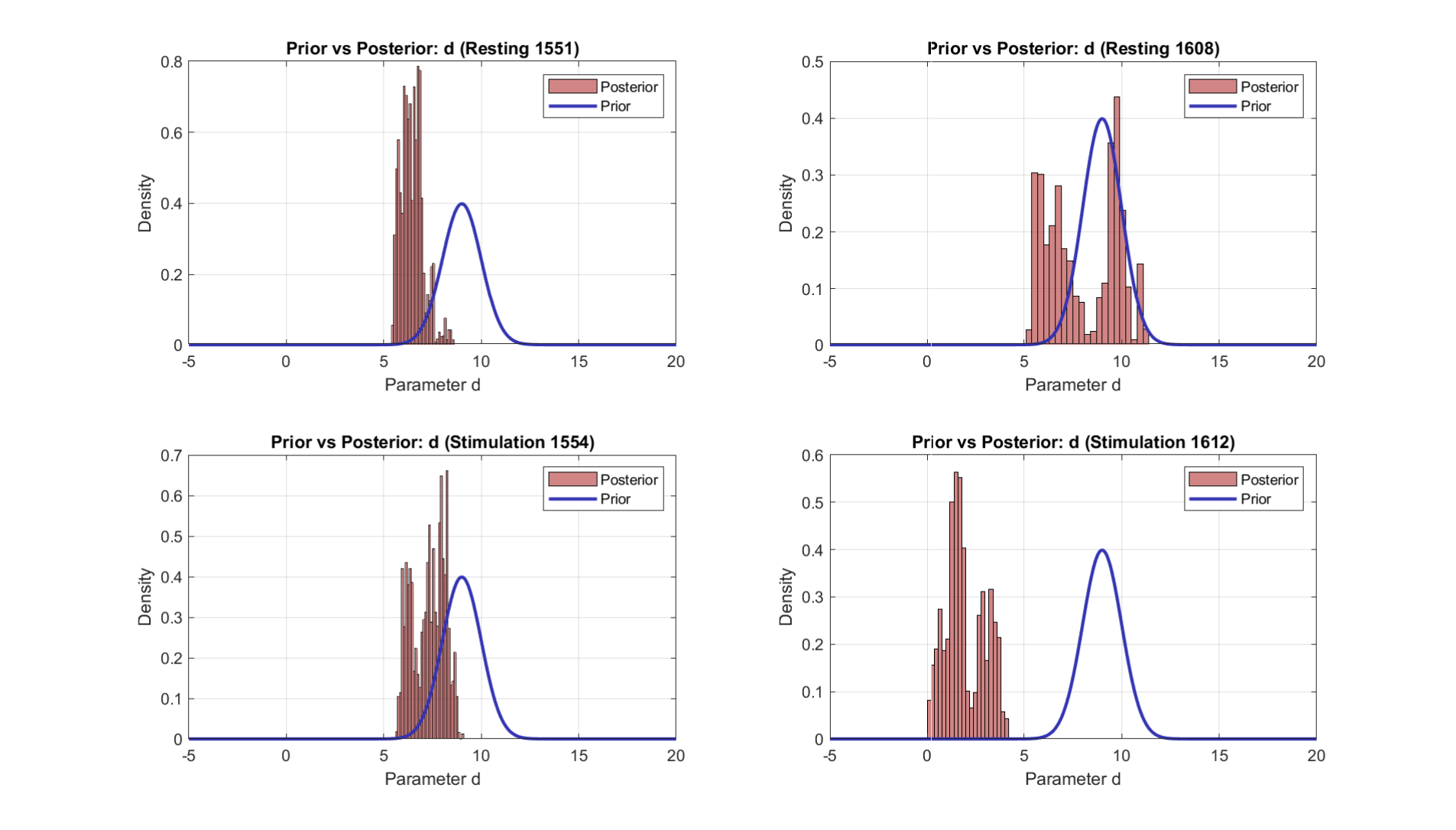}
    \caption{Posterior and Prior Distributions for Parameter \(d\) derived from Resting and Stimulated Real Data. The comparison indicates the adjustments in the reset recovery variable \(d\).}
    \label{fig:posteriord}
\end{figure}

\begin{figure}[H]
    \centering
    \includegraphics[width=0.8\textwidth]{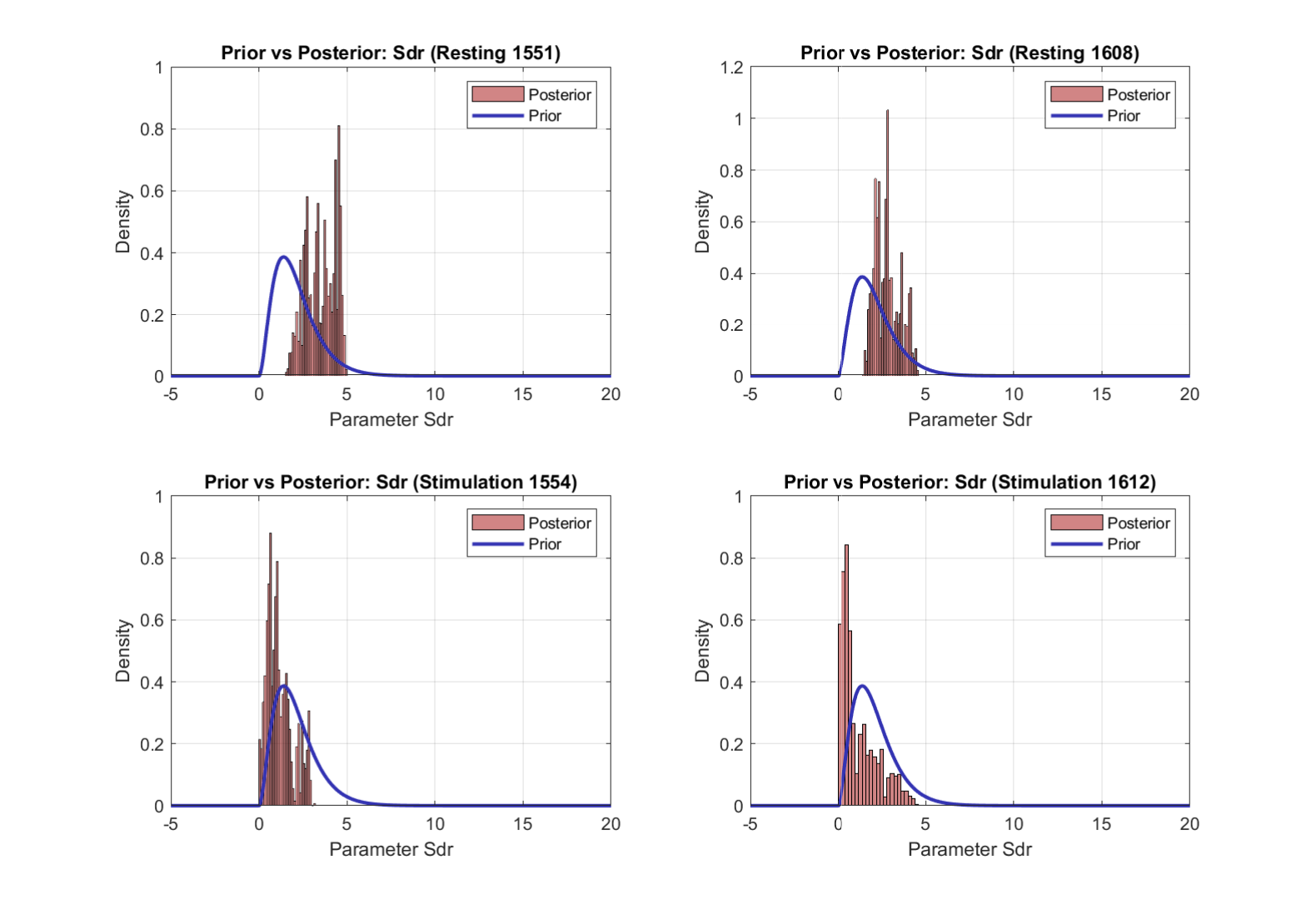}
    \caption{Posterior and Prior Distributions for Parameter \(S^{dr}\) derived from Resting and Stimulated Real Data. The posterior distribution reflects the estimated noise strength \(S^{dr}\).}
    \label{fig:posteriorSdr}
\end{figure}

The results from the PMMH analysis reveal significant differences in the posterior distributions of the parameters \(a\), \(b\), \(c\), \(d\), and \(S^{dr}\) between resting and stimulated cases. As shown in Figure \ref{fig:posteriora}, the posterior distribution of parameter \(a\), which governs the recovery time of the membrane potential, is more concentrated and lower in the resting state, indicating more consistent and efficient recovery during rest. Figure \ref{fig:posteriorb} highlights that parameter \(b\), related to recovery sensitivity, shifts towards higher values in the stimulated condition, reflecting increased neural responsiveness. For the reset membrane potential \(c\), displayed in Figure \ref{fig:posteriorc}, the posterior distribution remains centered near the prior for both conditions, but the stimulated state shows greater variability, suggesting dynamic resetting during neural activation. In Figure \ref{fig:posteriord}, the reset recovery variable \(d\) exhibits a posterior shift towards higher values under stimulation, signifying heightened recovery activity following spikes. Finally, Figure \ref{fig:posteriorSdr} demonstrates that the noise strength parameter \(S^{dr}\), which captures stochastic input effects, increases substantially during stimulation, underscoring the enhanced role of random processes in driving neural activity. These findings underscore the distinct parameter dynamics under resting and stimulated conditions, reflecting adaptive neural behavior across different states.

\section{\textcolor{black}{Discussion}} \label{sec:discussion}

\textcolor{black}{The brain represents a highly complex dynamical system, and understanding the underlying dynamics of membrane potential in both individual and interconnected neurons is fundamental for uncovering the mechanisms that drive action potential generation. Estimating model parameters from intracellular recordings presents significant challenges, particularly due to the complex stochastic nature of neuronal activity. Stochastic differential equations, driven by jump processes, provide an effective framework for modeling the evolution of membrane potentials in neurons. In this work, we employ Bayesian parameter inference to analyze both single-neuron dynamics and the interactions between excitatory and inhibitory cells, using the Leaky Integrate-and-Fire (LIF) model as a baseline. Our approach, which leverages synthetic observation data, proves to be an effective tool for estimating coupling parameters, providing valuable insights into the connectivity of neuronal networks. A central feature of our methodology is the use of Multilevel Markov Chain Monte Carlo methods -advanced numerical techniques that significantly improve the efficiency of statistical computations from dynamic simulations. These methods offer considerable promise for computational modelers in computational neuroscience, providing a robust approach to parameter inference in complex systems. Our numerical simulations have yielded important insights into parameter inference within this framework, and we prove methodology can be successfully extended to real-world neuroscience data. Specifically, we demonstrate the applicability of our approach using the Izhikevich model, which has been used extensively to simulate spiking neuron dynamics in response to real biological data. The Izhikevich model is particularly useful for modeling large-scale neural networks, and its incorporation into our framework allows for a more accurate estimation of coupling parameters in spiking networks with real data. This model's flexibility, especially in the context of experimental recordings from neural circuits, makes it an ideal candidate for studying real-world neural interactions. Furthermore, our approach is designed to minimize computational costs while maintaining a pre-specified level of accuracy, particularly in terms of mean square error. This efficiency is crucial in neuroscience research, where large datasets and complex models often pose substantial computational challenges. The estimation of parameters in spiking neural networks remains an open area of research, as highlighted by ongoing studies. By incorporating Multilevel Monte Carlo methods, we have made a meaningful contribution to inferring the coupling strengths between neurons in networks. This work represents a significant step toward understanding neuronal network interactions, with broad implications for both computational modeling and experimental neuroscience.}

\section{\textcolor{black}{Conclusion}} \label{sec:conclusion}

\textcolor{black}{In this paper we have shown first that a leaky integrate-and-fire model driven by jump processes can effectively serve as a substitute for the more intricate Bayesian neuron model, particularly in tasks related to inference within spiking neuronal networks. We have further adapted the multilevel Markov Chain Monte Carlo method by introducing a hierarchical approach to time discretization. This modification specifically aims to reduce the computational costs required to achieve a pre-specified mean square error. As a result, our approach contributes significantly to the advancement of efficient and accurate numerical simulations in neuroscience. In addition, we have extended our methodology to study the Izhikevich model with real neuroscience data, showcasing the applicability of our approach in the context of biological experiments. By combining the strengths of the Izhikevich model with the power of Bayesian inference and advanced numerical techniques, we provide new insights into neuronal network dynamics. This work paves the way for further exploration of spiking neuron models in real-world neuroscience, offering potential improvements in the understanding of neural connectivity and the dynamics underlying brain function.}


\end{document}